\documentclass
[peerreview]{IEEEtran}
\usepackage{hyperref} %
\hypersetup{
    colorlinks,
    linkcolor={blue!80!black},%
    citecolor={green!30!black},
    urlcolor={blue!80!black}
}

\usepackage{enumerate}

\usepackage{amsmath,amssymb,amsfonts}
\usepackage{bbm}
\usepackage{nicefrac}
\usepackage{amsthm}
\usepackage{algorithmic}
\usepackage{cleveref}
\usepackage{textcomp}
\def\BibTeX{{\rm B\kern-.05em{\sc i\kern-.025em b}\kern-.08em
T\kern-.1667em\lower.7ex\hbox{E}\kern-.125emX}}
\usepackage{physics}
\usepackage{siunitx}
\AtBeginDocument{\RenewCommandCopy\qty\SI}
\ExplSyntaxOn
\msg_redirect_name:nnn { siunitx } { physics-pkg } { none }
\ExplSyntaxOff
\usepackage{float}

\usepackage{graphicx}
\usepackage{xcolor}

\usepackage{comment}
\usepackage{amssymb} 
\usepackage{stmaryrd}
\renewcommand{\trace}{\mathrm{Tr}}
\AtBeginDocument{\RenewCommandCopy\qty\SI}

\theoremstyle{remark}	\newtheorem{theorem}{Theorem}
\theoremstyle{remark}	\newtheorem{lemma}[theorem]{Lemma}
\theoremstyle{remark}	\newtheorem{corollary}[theorem]{Corollary}
\theoremstyle{remark}	
\theoremstyle{remark} \newtheorem{definition}{Definition}
\theoremstyle{remark} \newtheorem{remark}{Remark}
\theoremstyle{remark} \newtheorem{example}{Example}

\newcommand{\identity}{\mathbbm{1}}

\renewcommand{\trace}{\mathrm{Tr}}

\newcommand{\markovC}[1]{%
\begin{tikzpicture}[#1]%
\draw (0,0.3ex) -- (1ex,0.3ex);%
\draw (0.5ex,0.3ex) circle (0.2ex);
\draw[white] (0.2ex,0) -- (0.5ex,0);%
\end{tikzpicture}%
}
\newcommand{\Cbar}{\markovC{scale=2}}

\usepackage[square,numbers,sort&compress]{natbib}
\usepackage[utf8]{inputenc}
\usepackage{mathrsfs}

\usepackage{accents}
\newlength{\dhatheight}

\usepackage{lscape}

\usepackage{tikz}
\usetikzlibrary{positioning}
\usetikzlibrary{decorations.markings}
\usetikzlibrary{arrows}
\usetikzlibrary{arrows.meta}
\usepackage[free-standing-units]{siunitx}
\usepackage[siunitx, RPvoltages]{circuitikz}

\makeatletter
\ctikzset{bipoles/my switch/height/.initial=.5}
\ctikzset{bipoles/my switch/width/.initial=.50}
\pgfcircdeclarebipole{}{}{myswitch}{%
\ctikzvalof{bipoles/my switch/height}}{\ctikzvalof{bipoles/my switch/width}}{
        \pgfsetlinewidth{\pgfkeysvalueof{/tikz/circuitikz/bipoles/thickness}\pgfstartlinewidth}
        \pgfpathmoveto{\pgfpoint{\pgf@circ@res@left}{0pt}}
        \pgfpathlineto{\pgfpoint{\pgf@circ@res@right}{.75\pgf@circ@res@up}}
        \pgfusepath{draw}
        \pgfsetlinewidth{\pgfstartlinewidth}        
        \pgftransformshift{\pgfpoint{\pgf@circ@res@left}{0pt}}
        \pgfnode{ocirc}{center}{}{}{\pgfusepath{draw}}
        \pgftransformshift{\pgfpoint{2\pgf@circ@res@right}{0pt}}
        \pgfnode{ocirc}{center}{}{}{\pgfusepath{draw}}
}
\def\pgf@circ@myswitch@path#1{\pgf@circ@bipole@path{myswitch}{#1}}
\compattikzset{my switch/.style = {\circuitikzbasekey, 
/tikz/to path=\pgf@circ@myswitch@path}}

\usepackage{xcolor}
\definecolor{mypurple}{rgb}{1,0,1}
	\definecolor{apricot}{rgb}{0.98, 0.81, 0.69}
		\definecolor{azure}{rgb}{0.0, 0.5, 1.0}
			\definecolor{darkmidnightblue}{rgb}{0.0, 0.2, 0.4}
				\definecolor{tearose}{rgb}{0.97, 0.51, 0.47}
					\definecolor{teagreen}{rgb}{0.82, 0.94, 0.75}
						\definecolor{indigo}{rgb}{0.29, 0.0, 0.51}
							\definecolor{tyrianpurple}{rgb}{0.4, 0.01, 0.24}

\tikzstyle{myedgestyle} = [-triangle 60]
\tikzstyle{block} = [draw, shape=rectangle, minimum height=3cm, minimum width=3cm, node distance=4cm, line width=0.5pt]
\tikzstyle{sum} = [draw, shape=circle, node distance=4cm, line width=0.5pt, minimum width=2.em]
\tikzstyle{mult} = [draw, shape=circle, node distance=3cm, line width=0.5pt, minimum width=2.em]
\tikzstyle{branch}=[fill,shape=circle,minimum size=4cm,inner sep=0pt]

\allowdisplaybreaks %

\title{Empirical Coordination of Quantum Correlations}
\author{%
    \IEEEauthorblockN{Husein Natur and Uzi Pereg} \\
    \IEEEauthorblockA{\normalsize Electrical and Computer Engineering and Helen Diller Quantum Center, Technion 
   }}

\begin{document}

\maketitle

\begin{abstract}
 We introduce the notion of empirical coordination for quantum correlations. %
 Quantum mechanics enables the calculation of probabilities for experimental outcomes, emphasizing statistical averages rather than detailed descriptions of individual events. 
This makes empirical coordination a natural and operationally meaningful framework for quantum systems—particularly in the context of nonlocal games, which rely on repeated measurements to assess performance.
We begin by analyzing networks with \emph{classical links}, focusing on the cascade network.
 For this setting, we establish the optimal coordination rates, which indicate the minimal resources required to simulate a quantum state on average.
    Providing the users with shared randomness, before communication begins, does not affect the optimal  rates for empirical coordination. 
    
    Our analysis starts with a basic two-node scenario and extends to cascade networks, including the special case of a network with an isolated node. 
    The results can be further generalized to other networks as our analysis includes a generic achievability scheme.
The optimal rate formula involves optimization over a collection of  state extensions.
This is a unique feature of the quantum setting, as  the classical parallel does not include optimization. 
As demonstrated through  examples,
 the performance depends heavily on the choice of decomposition.

We then extend the framework to
networks with \emph{quantum links},
focusing on a broadcast setting where the receivers have side information.
Finally, we discuss how our results provide new insights into the implementation and simulation of quantum nonlocal games in the empirical regime.
\end{abstract}

\begin{IEEEkeywords}
Quantum information theory, quantum communication, empirical coordination, quantum measurements, reverse Shannon theorem.
\end{IEEEkeywords}

\section{Introduction}
Shannon theory for point-to-point networks has had a profound influence on communication in the digital age \cite{soni2017mind,costello2007channel,arora2020survey}. 
However, the simplistic model of a single source-destination pair does not capture many critical aspects of real-world networks \cite{ElGamalKim:11b}. 
In practice, networked systems often involve multiple sources and destinations, requiring the network to compute functions or make decisions rather than merely transmit data. 
The Internet of Things (IoT) introduces additional challenges due to its reliance on a shared  medium \cite{he2020internet}. Furthermore, networks entail intricate tradeoffs between competition for resources \cite{1543455,825799}, cooperation for collective gain \cite{9706458}, and security \cite{10619085}.
Network information theory seeks to address fundamental questions of information flow and processing while incorporating these essential characteristics of real-world networks  \cite{tse1999linear,rosenberger2023identification,pereg2023multiple,8768393}.
Recent advances in IoT have drawn attention to the role of coordination in networks with diverse topologies \cite{torres2023message}.

Coordination is a fundamental framework in network information theory 
\cite{sudan2019communication}. %
Cuff et al. \cite{cuff2010coordination} introduced a general information-theoretic model for 
network coordination, where as opposed to traditional  coding tasks, 
 the objective is not to exchange messages between  network nodes %
 but rather  generate  correlation \cite{%
 le2017joint}.
Two types of coordination tasks were introduced 
in the classical framework
\cite{cuff2010coordination}. 
In strong coordination, the users 
produce actions in order to simulate a product distribution. That is, the \emph{joint distribution} resembles that of a particular memoryless 
source \cite{BlochKliewer:13c}. 
Empirical coordination imposes a weaker and less stringent condition compared to strong coordination.
It requires the type, i.e., the \emph{frequency} of 
 actions, to converge into a desired distribution \cite{mylonakis2020remote}. %
 There are many information-theoretic tasks that are closely related to coordination, such as 
channel/source simulation \cite{%
 berta2013entanglement,
 BennettDevetakHarrowShorWinter:14p,
wilde2018entanglement,%
state_generation_using_correlated_resource_2023,salehi2024quantum}, randomness extraction \cite{6670761,
 tahmasbi2020steganography},
 entanglement distribution \cite{10438882},
 state transformation \cite{george2023revisiting,george2024reexamination}, 
 state merging \cite{bjelakovic2013universal,horodecki2007quantum},
 entanglement dilution \cite{hayden2003communication,harrow2004tight,kumagai2013entanglement},
and
 compression \cite{goldfeld2014ahlswede,kramer2001quantum,Compressing_mixed_state_sources_2002, Classical_broadcast_cooperation_2016,
Quantum_Classical_Source_Coding2023,
Rate_limited_source_coding_2023}.

Empirical coordination %
and its variations are widely studied in the classical  information theory literature. %
Le Treust 
\cite{le2017joint} considered joint source-channel empirical coordination.
Le Treust and Bloch \cite{le2016empirical} further used empirical coordination as a unified perspective for
masking, amplification, and parameter estimation at the receiver.
Cuff and Zhao \cite{cuff2011coordination} studied 
empirical
coordination using implicit communication, with information embedding applications, such as  digital watermarking, steganography,
cooperative communication, and strategic play in team
games.
Cervia et al. \cite{cervia2016polar}
devised a polar coding scheme for empirical coordination.
Related models can also be found in 
\cite{blasco2014communication,haddadpour2012coordination}.
Quantum mechanics enables the calculation of probabilities for experimental outcomes, emphasizing statistical averages rather than detailed descriptions of individual events. For instance, the Heisenberg uncertainty principle states that the standard deviations of position and momentum cannot be minimized simultaneously \cite{fuchs1996quantum}. Some scholars, such as Fuchs and Peres \cite{fuchs2000quantum}, contend that quantum theory does not describe physical reality at all, but is instead confined to representing statistical correlations \cite{bricmont2016making}.
Empirical coordination is thus a natural framework for quantum systems.

Empirical coordination also plays a role in quantum data compression \cite{%
soljanin2002compressing}.
Barnum et al. \cite{barnum2001quantum} addressed a source of commuting density operators, and 
Kramer and Savari \cite{kramer2001quantum} developed a 
rate-distortion theory that unifies the visible and blind 
approaches (cf. \cite{dur2001visible} and \cite{horodecki1998limits}). Khanian and Winter have recently solved the general problem of a %
quantum source of mixed states %
(see also \cite{horodecki1998limits,horodecki1999towards,koashi2001compressibility,koashi2002operations,hayashi2006optimal,khanian2020quantum,khanian2022strong}). 

Coordination of quantum correlations with either classical or quantum links is described as follows. Consider a network of $K$ nodes, where Node $k$  performs an encoding operation $\mathcal{E}_k$ on a  system $A_k$, for $k\in \{1,\ldots,K\}$. Some of the nodes are connected by one-way classical links. %
We denote the classical rate and quantum rate limits for the link from Node $k$ to Node $l$ by $R_{k,l}$ and $Q_{k,l}$, respectively.
Before the coordination protocol begins, the nodes may also share %
common randomness (CR). Furthermore, some of the nodes can have access to side information. The objective in the coordination problem is to establish a specific correlation, i.e., to simulate %
a desired quantum state $\omega_{A_1\dots A_K}$. %
The optimal performance %
is defined by the  communication rates that are necessary and sufficient for simulating the desired correlation on average.

In analogy to the classical framework, we separate between
two types of coordination tasks. 
In strong coordination, the users 
encode in order to simulate an $n$-fold product state, $\omega_{A_1\dots A_K}^{\otimes n}$. That is, the \emph{joint state} resembles that of a  memoryless (i.i.d.) quantum source. 
In our previous work, we have considered strong coordination.
In particular, we addressed strong coordination for entanglement generation using quantum links
\cite{NaturPereg_E_arXiv,natur2025quantum} and 
for classical-quantum (c-q) correlations with classical links \cite{NaturPereg_CQ_ITW,natur2025quantum}. Strong coordination can be viewed as a unified framework for various models \cite{natur2025quantum}. We list a few examples of  related protocols:
\begin{enumerate}[1)]
    \item \emph{Channel resolvability:} 
Resolvability aims to approximate the output of a c-q channel using a uniformly distributed codebook %
\cite{hayashi2016quantum}.
This is equivalent to c-q state simulation. %
    \cite{%
    hayashi2016quantum}.
    Resolvability is also referred to as c-q soft covering \cite{strong_emprical_2018}. 
    Quantum soft covering is further studied in \cite{atif2023quantum}. %

    \item 
    \emph{Entanglement dilution and distillation:} In the  dilution task,  Alice and Bob use  a  maximally entangled state as well as local operations and classical communication (LOCC), in order to prepare a joint state 
    \cite{hayden2003communication,harrow2004tight}. 
    In the other direction, %
    maximal entanglement can be distilled from a bipartite state $\omega_{AB}$ %
    using %
    classical communication at a rate  $R_{1,2}\geq H(A|B)_\omega$ (see \cite{devetak2005distillation}). 
    A similar rate also appears in the distillation of  a secret key \cite[Remark~2]{devetak2005distillation}.
    Further work %
    can be found in \cite{christandl2007unifying,devetak2005distillation,bennett2014quantum,ekert1991quantum,%
    bennett1988privacy,dupuis2023privacy,shen2024optimal,%
    berta2013quantum,cheng2016randomness,anco2024much}. %

    \item 
    \emph{State merging and splitting:} %
    In state merging, Alice and Bob share $\omega_{AB}$, and
    Alice would like to 
    send her part %
    to Bob 
    \cite{horodecki2005partial,horodecki2007quantum}.
       The mother protocol generalizes this task \cite{abeyesinghe2009mother,berta2016smooth}.
Whereas, state splitting 
is the reverse task, where 
Alice holds $AB$, and would like to send $B$ to Bob \cite{devetak2005triangle,oppenheim2008state,berta2011quantum}.
   \item 
    \emph{Channel simulation:} %
    A classical channel of capacity $C$ can be simulated at a rate of $R_{1,2}$ if and only if $R_{1,2}\geq C$, given sufficient %
 common randomness \cite{bennett2002entanglement,cuff2008communication}. %
    The quantum analog is not  necessarily true %
        \cite{bennett2014quantum}.
    The %
    entanglement cost with LOCC
    is related to %
    the entanglement of formation
    \cite{berta2013entanglement}.
\end{enumerate}

Multi-user versions of the protocols above have  been studied extensively in recent years. 
The mother protocol can generate
 distributed compression protocols for correlated quantum sources %
 \cite{abeyesinghe2009mother,ahn2006distributed,KhanianWinter:18a,salek2018quantum,Faithful_simulation_Heidari_2019,%
 khanian2024rate,colomer2024decoupling}.
Simulation of broadcast and multiple-access channels is considered in \cite{cheng2023quantum,cao2024channel} and 
 \cite{nema2024one}, respectively. 
  George and Cheng~\cite{george2024coherent} have recently studied multipartite state splitting. 
    Multi-user distillation and manipulation is considered in  \cite{smolin2005entanglement,bravyi2006ghz,augusiak2009multipartite,streltsov2017rates,murta2020quantum,Salek_2022_Winter,salek2023new}.
 Streltsov et al. \cite{streltsov2020rates} studied 
 multipartite state merging.  
A more detailed overview is given in \cite{NP_2024}.
Here, we introduce the notion of 
\emph{empirical coordination} for quantum correlations, 
imposing a weaker and less stringent condition compared to strong coordination.
We require the \emph{empirical average} state 
 to converge into the desired state $\omega_{A_1\dots A_K}$.
Specifically, let $\mathbf{A}(1),\ldots,\mathbf{A}(n)$ denote the output sequence from all network nodes, where $\mathbf{A}(i)\equiv (A_1(i),\ldots,A_K(i))$ is the output, at time $i$, for $i\in  \{{1,  \dots , n}\}$.
 Then, we would like the nodes to produce an empirical average state 
 $\frac{1}{n}\sum_{i=1}^n \rho_{\mathbf{A}(i)}$
 that is arbitrarily close to $\omega_{\mathbf{A}}$, where $\mathbf{A}\equiv (A_1,\ldots,A_K)$.
 That is, we require that the distance,
 \begin{align}
\norm{ \frac{1}{n}\sum_{i=1}^n \rho_{\mathbf{A}(i)} -\omega_{\mathbf{A}} }_1
 \end{align}
converges to zero as the block length $n$ tends to infinity. %
Our networks consist of nodes possessing quantum systems, and are connected with either classical or quantum links of limited communication rates.

\begin{figure}[tb]
\center
\includegraphics[scale=0.75,trim={5.3cm 0 5.5cm 0}]
{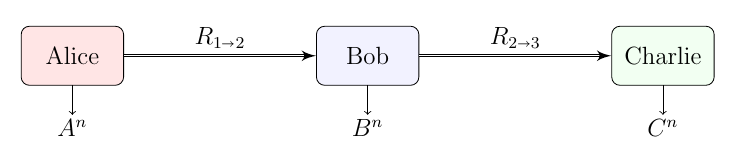} %
\caption{Cascade network with classical links.
}
\label{Figure: Introduction cascade network}
\end{figure}

\begin{figure}[tb]
\center
\includegraphics[scale=0.75,trim={5.3cm 0 5.5cm 0}]
{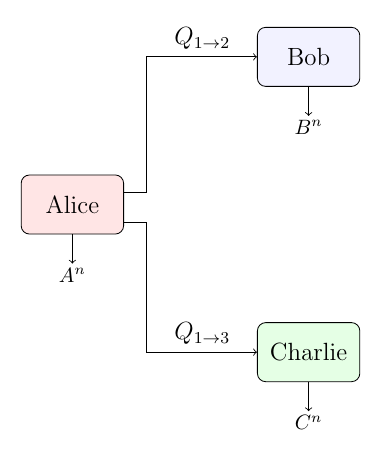} %
\caption{Broadcast network with quantum links.
}
\label{Figure: Broadcast network with quantum links - Introduction}
\end{figure}

After introducing the definition of empirical coordination for quantum states, we
 discuss the justification for our  definition   and its physical interpretation. We focus on the 3-node cascade network and broadcast and determine the optimal coordination rates,  
which represent the minimal resources  for the empirical simulation of a quantum state among multiple parties. 
Theses networks with $K=3$ users can be viewed as a building block for larger multiuser systems \cite{9133575,jiang2024quantum}.
The  cascade setting is depicted
 in Figure~\ref{Figure: Introduction cascade network}. Alice, Bob, and Charlie wish to simulate a separable state $\omega_{ABC}$.
 They are provided with rate-limited classical communication links, %
   $R_{1 \shortrightarrow 2}$ from Alice to Bob, and $R_{2 \shortrightarrow 3}$ from Bob to Charlie.
   Then, we consider a broadcast network with quantum links, as in Figure~\ref{Figure: Broadcast network with quantum links - Introduction}, where Alice distributes entanglement to two receivers, Bob and Charlie, at the respective quantum rates $Q_{1\to 2}$ and $Q_{1\to 3}$.

   Our results are summarized below. We consider empirical coordination in two types of networks, with classical links and quantum links. For networks with classical communication links,  
   we show that CR between the network users does not affect the optimal  rates for empirical coordination. We begin with the rate characterization for the basic two-node network, and then generalize to a cascade network.  The special case of a network with an isolated node is considered as well.  The results can be further generalized to other networks as our analysis includes a generic achievability scheme.
The characterization involves optimization over a collection of  state extensions.
This is a unique feature of the quantum setting, as  the classical parallel does not include optimization \cite{cuff2010coordination}. 
As will be seen in the examples,
 the performance depends heavily on the choice of decomposition.
 We further discuss the consequences of our results for  cooperative games. 
 
 We then present our results on networks with quantum links. We begin with the rate characterization for the basic two-node network with a quantum link, and then determine the optimal rates in the broadcast network, %
 where Alice distributes qubits to Bob and Charlie, thus creating tripartite entanglement between Alice, Bob, and Charlie. %
 We then demonstrate a tight connection between our results and the theory of nonlocal correlations and refereed games with quantum strategies, highlighting the implications of this relationship on the statistiscs collected in a game realization.
 
 In Section~\ref{Section: Notation}, we set our notation conventions. In Section~\ref{Section: Model definitions}, we present the definitions for our model and their physical interpretation.
 Section~\ref{Section: CL Results} is dedicated to the results for networks with classical links, including the statement about CR and the capacity theorems for the two-node network, the cascade network and the isolated node. In Section~\ref{Section: QL Results}, we present our results on networks  %
 with quantum links. %
 In Section~\ref{Section:Nonlocal_games}, we discuss the implications of our results on quantum nonlocal games. 
 Section~\ref{Section: Achievability analysis} and Section~\ref{Section: Converse analysis} provide the achievability and converse analysis, respectively, for networks with classical links, and Section~\ref{Section: Analysis - Quantum links} for quantum links. %
 Section~\ref{Section: Discussion} concludes with a discussion on the comparison between strong coordination and empirical coordination, as well as the implications of our results on quantum cooperative games.


\section{Notation}
\label{Section: Notation}
We use standard notation in quantum information theory,  as in
\cite{wilde2017quantum},
$X,Y,Z,\ldots$ are discrete random variables on finite alphabets   $\mathcal{X},\mathcal{Y},\mathcal{Z},...$, respectively, 
  The distribution of  $X$ is specified by a probability mass function (PMF) $p_X(x)$ on $\mathcal{X}$. 
  The set of all 
  PMFs over $\mathcal{X}$ is denoted by $\mathcal{P}(\mathcal{X})$. The normalized total variation distance between two PMFs in $\mathcal{P}(\mathcal{X})$ is defined as
\begin{align}
\frac{1}{2}\norm{p_X-q_X}_1 =\frac{1}{2}\sum_{a\in \mathcal{X}}\abs{p_X(a)-q_X(a)} 
\end{align}
for every $p_X$, $q_X\in$  $\mathcal{P}(\mathcal{X})$.

  The classical Shannon entropy is then defined as $H(p_X)= \sum_{x\in\mathrm{supp}(p_X)} p_X(x)\log\left( \frac{1}{p_X(x)} \right) $, with logarithm to base $2$.
  We often use the short notation 
  $H(X)\equiv H(p_X)$ for 
  $X\sim p_X$.
  Similarly, 
given a joint PMF $p_{XY}\in\mathcal{P}(\mathcal{X}\times\mathcal{Y})$, we write $H(XY)\equiv H(p_{XY})$.
The mutual information between $X$ and $Y$ is 
$I(X;Y)= H(X)+H(Y)-H(XY)$.
A classical channel is defined by a probability kernel  
$\{
p_{Y|X}(y|x)
\,:\;
x\in\mathcal{X}, y\in\mathcal{Y}
\}$. The conditional entropy with respect to 
$p_X\times p_{Y|X}$
is defined as
$H(Y|X)=\sum_{x\in\mathrm{supp}(p_X)} p_X(x)H(Y|X=x)
$, where $H(Y|X=x)\equiv H(p_{Y|X}(\cdot|x))$.
According to the entropy chain rule, $H(Y|X)=H(XY)-H(X)$.

We use 
$x^n=(x_i:\, i\in [n])$ for  
a sequence  of letters from  
 $\mathcal{X}$, where $[n]\equiv \{{1,  \dots , n}\}$.
We define the type of a sequence $x^n$ as the empirical distribution, $\hat{P}_{x^n}(a)=\frac{1}{n}N(a|x^n)$, where $N(a|x^n)$ is the number of occurrences of the letter $a$ in the sequence $x^n$, for $a\in\mathcal{X}$. 
The $\delta$-typical set for a PMF $p_X$,
is defined here as
 the set of sequences whose type is $\delta$-close to $p_X$ in total variation distance. Formally,
\begin{align}
T_\delta^{(n)}(p_X)\equiv \left\{x^n\in \mathcal{X}^n:\frac{1}{2}\norm{\hat{P}_{x^n}-p_X}_1 < \delta \right\} 
\,.
\end{align}

A quantum system is associated with a Hilbert space, 
$\mathcal{H}$. 
The dimensions are assumed to be finite throughout. Denote the set of all linear operators $F:\mathcal{H}\to \mathcal{H}$ by $L(\mathcal{H})$.
The Hermitian conjugate of $F$ is denoted $F^\dagger$.
The extension of a real-valued function to Hermitian operators is defined in the usual manner.
Analogically to the total variation distance between  classical PMFs, the normalized trace distance between two Hermitian operators  satisfies 
\begin{align}
        \frac{1}{2}\norm{P-Q}_{1}=\frac{1}{2}\Tr\Big[\abs{P-Q}%
        \Big]
\end{align}
for every Hermitian
$P$, $Q\in$ $L(\mathcal{H})$. 

Let \emph{System $A$} be associated with $\mathcal{H}_A$.
The quantum state of $A$ is described by a density operator $\rho_A\in L(\mathcal{H}_A)$, i.e., a unit-trace positive semidefinite operator.
Let $\Delta(\mathcal{H}_A)$
denote the set of all such %
operators.
The probability distribution of a  measurement outcome  is derived from a positive operator-valued measure (POVM). In finite dimensions, this reduces to a finite set of positive semidefinite operators $\{D_j \,:\; j\in [N] \}$ that satisfy  
$\sum_{j=1}^N D_j=\identity$, where $\identity$ denotes the identity operator.
By the Born rule, 
the probability of a measurement outcome $j$ is given by $p_J(j)=\trace(D_j \rho_A)$, for $j\in [N]$.

The von Neumann entropy of a quantum state $\rho_A\in\Delta(\mathcal{H}_A)$ is defined as 
$H(\rho_A) \equiv -\trace[ \rho_A\log(\rho_A) ]$. We often denote the quantum entropy by  
$H(A)_\rho\equiv H(\rho_A)$.
Similarly,
given a joint state $\rho_{AB}\in \Delta(\mathcal{H}_A\otimes \mathcal{H}_B)$,  
we write $H(AB)_\rho\equiv H(\rho_{AB})$.
A pure state has zero entropy, in which case, there exists $\ket{\psi}\in\mathcal{H}_A$ such that $\rho=\ketbra{\psi}$, where 
$\bra{\psi}\equiv (\ket{\psi})^\dagger$.
The conditional quantum entropy is  \emph{defined} by 
$H(A|B)_{\rho}=H(AB)_\rho-H(B)_\rho$. The conditional mutual information is defined accordingly, as
$I(A;B|C)_{\rho}\equiv H(A|C)_\rho+H(B|C)_\rho-H(A,B|C)_\rho$ for 
$\rho_{ABC}\in\Delta(\mathcal{H}_A\otimes \mathcal{H}_B\otimes \mathcal{H}_C)$. %

 A bipartite state $\rho_{AB}$  is said to be \emph{separable} if a set of product states 
 $\{\rho_x\otimes \sigma_x\}$
 in $\Delta(\mathcal{H}_A\otimes \mathcal{H}_B)$ can be found such that 
 \begin{align}
\rho_{AB}=
\sum_{x\in\mathcal{X}} p_X(x)
\rho_x\otimes \sigma_x
 \end{align}
 for some alphabet $\mathcal{X}$ and
 PMF $p_X$ on  $\mathcal{X}$.
 Otherwise, $\rho_{AB}$ is called \emph{entangled}.
 If the state is entangled, then
 the conditional entropy $H(A|B)_\rho$  can be negative. 
 The definition can also be extended to a multipartite system. A  state $\rho_{A_1\ldots A_K}$ in $\Delta(\mathcal{H}_{A_1}\otimes\cdots\otimes \mathcal{H}_{A_K})$  is said to be \emph{separable} if  
 \begin{align}
\rho_{A_1\ldots A_K}=
\sum_{x\in\mathcal{X}} p_X(x)
\rho_x^{(1)}\otimes\cdots\otimes \rho^{(K)}_x
 \end{align}
 for some ensemble  $\{p_X\,,\; \rho^{(1)}_x\otimes\cdots\otimes \rho^{(K)}_x \,,\; x\in\mathcal{X}\}$.

A quantum channel is defined by a completely-positive trace-preserving map, $\mathcal{N}_{A\to B}: L(\mathcal{H}_A)\to L(\mathcal{H}_B)$. In general, the channel maps a state $\rho\in \Delta(\mathcal{H}_A)$ into a state 
$\mathcal{N}_{A\to B}(\rho)\in \Delta(\mathcal{H}_B) $.
A classical-quantum (c-q) channel  $\mathcal{N}_{X\to B}$
is specified by a collection of quantum states $\{\rho_B^{(x)}\,:\;
x\in\mathcal{X}
\}$ in $\Delta(\mathcal{H}_B) $, where
$\rho_B^{(x)}\equiv \mathcal{N}_{X\to B}(x)$ for $x\in\mathcal{X}$.

\section{Model Definition and Physical Interpretation}%
\label{Section: Model definitions}

\begin{figure}[tb]
\center
\includegraphics[scale=0.75, trim={5.3cm 0 5.5cm 0}]{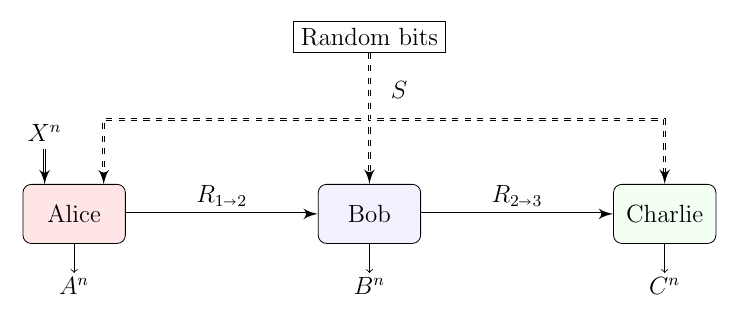}
\caption{Cascade network with classical links and common randomness.
}
\label{Figure: Cascade network with CR}
\end{figure}

\subsection{Coding definitions}
In this subsection, we introduce the basic definitions for empirical coordination. 
Consider the cascade network shown in Figure~\ref{Figure: Cascade network with CR}, which involves three users, Alice, Bob, and Charlie.
Let $\{ p_X(x)\,,\; \omega_{ABC}^x \,,\; x\in\mathcal{X} \}$ be a given ensemble, with an average
\begin{align}
\omega_{ABC}=
\sum_{x\in\mathcal{X}}
p_X(x) \omega_{ABC}^x
\,.
\end{align}
Suppose that
Alice receives a random sequence $X^n$, drawn from a memoryless (i.i.d.) source $\sim p_X$.
This can be viewed as side information  Alice obtains from a local measurement on her environment.
Alice sends a classical message 
$m_{1 \shortrightarrow 2}$ to Bob via a noiseless link of limited rate $R_{1 \shortrightarrow 2}$, and Bob sends  $m_{2 \shortrightarrow 3}$ to Charlie at a limited rate $R_{2 \shortrightarrow 3}$. %
Next, Alice, Bob, and Charlie encode their respective quantum outputs  $A^n,B^n$, and $C^n$. The objective of the empirical coordination protocol is for the average  state  to be arbitrarily close %
to a particular %
state $\omega_{ABC}$.

\begin{remark}
Achieving empirical coordination  allows the network users to perform local measurements such that the outcome statistics follow a desired behavior.
\end{remark}
In other words, the users utilize a  coding scheme that simulates on average a desired state $\omega_{ABC}$.
We are interested in the lowest communication rates
$(R_{1 \shortrightarrow 2},R_{2 \shortrightarrow 3})$ 
that are required in order to achieve this goal.  

In the beginning, we assume that Alice, Bob, and Charlie share unlimited common randomness (CR). That is, a random element $S$ is drawn a priori and distributed to Alice, Bob, and Charlie before the protocol begins.
Later, we will show that CR does not affect the achievable rates.

\begin{definition}
    A $(2^{nR_{1 \shortrightarrow 2}},2^{nR_{2 \shortrightarrow 3}},n)$ empirical coordination code for the cascade network shown in Figure~\ref{Figure: Cascade network with CR} consists of: 
    \begin{itemize}
    \item
    a CR source $p_S$ over  a randomization set $\mathfrak{S}_n$, 
    \item
    a pair of classical encoding channels,
    $p_{M_{1\shortrightarrow 2}|X^n S}$ and $ p_{M_{2 \shortrightarrow 3}|M_{1 \shortrightarrow 2}S}$, over the index sets %
 $ \big[2^{nR_{1 \shortrightarrow 2}}\big]$ and $\big[2^{nR_{2 \shortrightarrow 3}}\big]$, respectively, and
 
 \item
 c-q encoding channels, \begin{align}
 &
\mathcal{E}_{XS\to A}: \mathcal{X}\times \mathfrak{S}_n \to 
\Delta(\mathcal{H}_A) 
\,,\\ 
&
     \mathcal{F}_{M_{1 \shortrightarrow 2} M_{2 \shortrightarrow 3} S\to B^n }: [2^{n R_{1 \shortrightarrow 2}}] \times[2^{n R_{2 \shortrightarrow 3}}] \times \mathfrak{S}_n \to \Delta(\mathcal{H}_B^{\otimes n}) \,, 
\intertext{and}
&
    \mathcal{D}_{M_{2 \shortrightarrow 3}S\to C^n}: [2^{n R_{2 \shortrightarrow 3}}] \times \mathfrak{S}_n \to \Delta({\mathcal{H}_C}^{\otimes n})\,,
\end{align}
for Alice, Bob, and Charlie, respectively,
where $\mathfrak{S}_n$ is an unbounded set of realizations for the CR resource that is shared between the users a priori.
\end{itemize}
\end{definition}

The %
protocol works as follows.
Before communication begins, 
Alice, Bob, and Charlie share a CR element $s$, drawn from the source $ p_S$.
Alice receives a sequence $x^n$, generated from 
 a memoryless source $p_X$. That is, the random sequence is distributed according to $p_X^n(x^n)\equiv \prod_{i=1}^n p_X(x_i)$.
She selects an index
\begin{align}
m_{1 \shortrightarrow 2}\sim
p_{M_{1 \shortrightarrow 2}|X^n S}(\cdot|x^n,s) %
\end{align}
at random,
and sends it through a noiseless classical link at rate $R_{1 \shortrightarrow 2}$.
She then applies the encoding channel $\mathcal{E}_{X S\to A}^{\otimes n}$, to prepare the state of her system $A^n$, hence
\begin{align}
        \rho_{A^n}^{(x^n,s)}&=\bigotimes_{i=1}^n\mathcal{E}_{X S\to A}(x_i,s)\,.
        \label{Equation: Alice Cascade states with CR}
    \end{align}

As Bob receives the message $m_{1 \shortrightarrow 2}$ and the CR element $s$, he selects a random index 
\begin{align}
m_{2 \shortrightarrow 3}\sim p_{M_{2 \shortrightarrow 3}|M_{1 \shortrightarrow 2}S}(\cdot|m_{1 \shortrightarrow 2},s) %
\end{align}
and sends it through a noiseless classical link at rate $R_{2 \shortrightarrow 3}$ to Charlie.
Bob and Charlie encode their systems, $B^n$ and $C^n$, by 
\begin{align}
        \rho_{B^n}^{(m_{1 \shortrightarrow 2},m_{2 \shortrightarrow 3},s)}&=\mathcal{F}_{M_{1 \shortrightarrow 2}M_{2 \shortrightarrow 3}S\to B^n}(m_{1 \shortrightarrow 2},m_{2 \shortrightarrow 3},s)\,,
        \label{Equation: Bob Cascade states with CR}
\intertext{and}\rho_{C^n}^{(m_{2 \shortrightarrow 3},s)}&=\mathcal{D}_{M_{2 \shortrightarrow 3}S\to C^n}(m_{2 \shortrightarrow 3},s)%
  \label{Equation: Charlies Cascade states with CR}
    \end{align}
    respectively.
    
Given a value $s$, i.e., a realization  of the random element,   consider 
the average state 
$\overline{\rho}_{ABC}(s)\in
\Delta(\mathcal{H}_A\otimes
\mathcal{H}_B\otimes
\mathcal{H}_C)$
that is induced by the code: 
    \begin{multline}
        \overline{\rho}_{ABC}(s)\equiv\frac{1}{n}\sum_{i=1}^{n}
        \sum_{x^n\in\mathcal{X}^n} 
        \sum_{m_{1 \shortrightarrow 2}\in[2^{n{R_{1 \shortrightarrow 2}}}]}\sum_{m_{2 \shortrightarrow 3}\in[2^{n{R_{2 \shortrightarrow 3}}}]} p_X^n(x^n) p_{M_{1 \shortrightarrow 2}|X^n S}(m_{1 \shortrightarrow 2}|x^n,s) p_{M_{2 \shortrightarrow 3}|M_{1 \shortrightarrow 2} S}(m_{2 \shortrightarrow 3}|m_{1 \shortrightarrow 2},s)
        \\
        \cdot
        \rho_{A_i}^{(x_i,s)}\otimes \rho_{B_i}^{(m_{1 \shortrightarrow 2},m_{2 \shortrightarrow 3},s)} \otimes \rho_{C_i}^{(m_{2 \shortrightarrow 3},s)}
        \,.
    \end{multline}
We now define achievable rates as rates that are sufficient in order to encode $\overline{\rho}_{ABC}(s)$ that converges to $\omega_{ABC}$.

\begin{definition}
\label{Definition: Achievable rate pair - Cascade with CR}
    A rate pair $\left(R_{1 \shortrightarrow 2},R_{2 \shortrightarrow 3}\right)$ is achievable for empirical coordination of a desired separable state $\omega_{ABC}$,
    if for every $\alpha,\varepsilon,\delta>0$ and a sufficiently large $n$, there exists a  $\left(2^{n(R_{1 \shortrightarrow 2}+\alpha)},2^{n(R_{2 \shortrightarrow 3}+\alpha)},n\right)$ 
coordination 
code  that achieves
\begin{align}
\Pr\left(\frac{1}{2}\norm{\overline{\rho}_{ABC}(S)-\omega_{ABC}}_1 > \varepsilon \right) \leq \delta  \,,
\end{align}
where the probability is computed with respect to the CR element $S\sim p_S$.

Equivalently, there exists a sequence of empirical coordination codes such that the error converges to zero in probability, i.e.,
\begin{align}
\norm{\overline{\rho}_{ABC}(S)-\omega_{ABC}}_1\longrightarrow 0 \; \text{in probability.}
\end{align}
\end{definition}

For a network with quantum links, we define an empirical coordination code
in a similar manner, where the classical registers of $M_{k,l}$ are replaced by quantum systems, which are sent from Node $k$ to Node $l$ through a noiseless quantum link.  The link has a quantum rate $Q_{k,l}$.
Further details will be given in Section~\ref{Section: QL Results}.

\begin{remark}
\label{Remark: Separable}
In Section~\ref{Section: CL Results}, we focus on networks with classical 
 communication links, where
 entanglement cannot be generated.
 Therefore, we only consider separable states $\omega_{ABC}$ for such networks.
 Later, in Section~\ref{Section: QL Results}, we consider empirical coordination of entanglement in networks with quantum links.
\end{remark}

\subsection{Quantum measurements}
\label{Subsection: Justification}
In this subsection, we discuss the justification and the physical interpretation of our coordination criterion. 
As mentioned in the Introduction, quantum mechanics enables the calculation of probabilities for experimental outcomes, emphasizing statistical averages rather than detailed descriptions of individual events. For instance, the Heisenberg uncertainty principle states that the standard deviations of position and momentum cannot  be minimized simultaneously \cite{fuchs1996quantum}.
Some scholars, such as Fuchs and Peres \cite{fuchs2000quantum}, contend that quantum theory does not describe physical reality at all, but is instead confined to representing statistical correlations \cite{bricmont2016making}.
Empirical coordination is thus a natural framework for quantum systems.
Further justification is provided below. 

Consider an observable represented by an Hermitian operator $\hat{O}$ on $\mathcal{H}_A\otimes \mathcal{H}_B\otimes \mathcal{H}_C$. In practice, statistics are collected by performing measurements on $n$ systems
$(A_i,B_i,C_i \,:\; i\in [n])$.
The expected value of the observable in the $i$th measurement is thus,
\begin{align}
\langle \hat{O} \rangle_i&=
\trace\left[ \hat{O}\cdot \rho_{A_i B_i C_i} \right]
\end{align}
for $i\in [n]$.
Therefore, the empirical average  is
\begin{align}
\frac{1}{n}\sum_{i=1}^n\langle \hat{O} \rangle_i&=
\trace\left[ \hat{O} \cdot \left( \frac{1}{n}\sum_{i=1}^n \rho_{A_i B_i C_i} \right) \right]
\nonumber\\
&=
\trace\left[ \hat{O}\cdot   \overline{\rho}_{A B C}  \right]
\end{align}

Similarly, consider a POVM $\{D_\ell: \ell\in [L]\}$ on  $\mathcal{H}_A\otimes \mathcal{H}_B\otimes \mathcal{H}_C$. %
 The probability that we obtain the measurement outcome $\ell$ in the  $i$th measurement is 
$%
p_i(\ell)=\trace(D_\ell \cdot \rho_{A_i B_i C_i})
$. %
Thereby, the average distribution is  given by 
\begin{align}
\bar{p}(\ell)=\trace(D_\ell \cdot \overline{\rho}_{A B C})
\,.
\end{align}

\section{Main Results - Classical Links}
\label{Section: CL Results}
\subsection{Common randomness does not help}%

\begin{theorem}
\label{Theorem: CR theorem}
    Any desired %
    state $\omega_{ABC}$ that can be simulated at rate %
    $(R_{1 \shortrightarrow 2},R_{2 \shortrightarrow 3})$ through %
    empirical coordination in the cascade network  with CR assistance, can also be simulated with no CR, i.e., with $|\mathfrak{S}_n|=1$.
\end{theorem}
We will discuss the interpretation of this result in Subsection~\ref{Section: CR}.
 The proof %
 is provided below. 
\begin{proof}
Let  $(R_{1 \shortrightarrow 2},R_{2 \shortrightarrow 3})$ be an achievable rate pair for empirical coordination. %
    Consider the setting %
    in %
    Section~\ref{Section: Model definitions}.
   Let the CR element $S$ and the classical side information $X^n$ be drawn according to $p_S$ and $p_X^n$, respectively.
   Then,  
    Alice, Bob, and Charlie encode by 
\begin{align}
        &m_{1 \shortrightarrow 2}\sim p_{M_{1 \shortrightarrow 2}|X^nS}(\cdot|x^n,s)
        \,,\;
        \rho_{A^n}^{(x^n,s)}=\bigotimes_{i=1}^n\mathcal{E}_{XS\to A}(x_i,s)\,,
        \\
       & m_{2 \shortrightarrow 3}\sim p_{M_{2 \shortrightarrow 3}|M_{1 \shortrightarrow 2}S}(\cdot|m_{1 \shortrightarrow 2},s)
        \,,\;
        \rho_{B^n}^{(m_{1 \shortrightarrow 2},m_{2 \shortrightarrow 3},s)}=\mathcal{F}_{M_{1 \shortrightarrow 2}M_{2 \shortrightarrow 3} S\to B^n}(m_{1 \shortrightarrow 2},m_{2 \shortrightarrow 3},s)\,,
        \\
        &\rho_{C^n}^{(m_{2 \shortrightarrow 3},s)}=\mathcal{D}_{M_{2 \shortrightarrow 3}S\to C^n}(m_{2 \shortrightarrow 3},s)\,.
\end{align}
Denote the normalized trace distance by 
\begin{align}
    d(s)=\frac{1}{2} \norm{\overline{\rho}_{ABC}(s)-\omega_{ABC}}_1
\end{align}
for $s\in\mathfrak{S}_n$.

According to Definition~\ref{Definition: Achievable rate pair - Cascade with CR}, if a rate pair $(R_{1 \shortrightarrow 2},R_{2 \shortrightarrow 3})$ is achievable, then for every $\alpha, \varepsilon,\delta>0$ and sufficiently large $n$, there exists a sequence  of  $\left(2^{n(R_{1 \shortrightarrow 2}+\alpha)},2^{n(R_{2 \shortrightarrow 3}+\alpha)},n\right)$ empirical coordination codes, for which the following holds:
\begin{align}
\Pr\left(d(S)%
> \frac{\varepsilon}{2} \right) \leq \delta  \,.
\label{Equation: CR probability bound}
\end{align}
Averaging over the CR element yields the following average state
\begin{align}
    \hat{\rho}_{A^nB^nC^n} &= \mathbb{E}\left[\rho_{A^n}^{(m_{1 \shortrightarrow 2},S)}\otimes \rho_{B^n}^{(m_{1 \shortrightarrow 2},S)} \otimes \rho_{C^n}^{(m_{2 \shortrightarrow 3},S)}\right]
    \nonumber \\
    &= \sum_{s\in \mathfrak{S}_n} p_S(s)\rho_{A^n}^{(m_{1 \shortrightarrow 2},s)}\otimes \rho_{B^n}^{(m_{1 \shortrightarrow 2},s)} \otimes \rho_{C^n}^{(m_{2 \shortrightarrow 3},s)}
    \,.
\end{align}

By the total expectation formula, 
\begin{align}
    \mathbb{E}\left[d(S)\right]&=\Pr\left(d(S)>\frac{\varepsilon}{2}\right)\cdot \mathbb{E}\left[d(S)\big|d(S)>\frac{\varepsilon}{2}\right]+\Pr\left(d(S)\leq\frac{\varepsilon}{2}\right)\cdot \mathbb{E}\left[d(S)\big|d(S)\leq\frac{\varepsilon}{2}\right]
    \nonumber\\
    &\leq \delta \cdot 1 + 1 \cdot \frac{\varepsilon}{2}
    \nonumber\\
    & <\varepsilon
\end{align}
where the second line follows from \eqref{Equation: CR probability bound}, and the last inequality holds by choosing $\delta<\frac{\varepsilon}{2}$.
Therefore, there exists  $s^*\in\mathfrak{S}_n$ for which $d(s^*)\leq \varepsilon$.
We can thus 
satisfy the coordination requirement with the following encoding maps, 
\begin{align}
        &m_{1 \shortrightarrow 2}\sim p_{M_{1 \shortrightarrow 2}|X^nS}(\cdot|x^n,s^*)
        \,,\;
        \rho_{A^n}^{(x^n)}=\bigotimes_{i=1}^n\mathcal{E}_{XS\to A}(x_i,s^*)\,,
        \\
       & m_{2 \shortrightarrow 3}\sim p_{M_{2 \shortrightarrow 3}|M_{1 \shortrightarrow 2}S}(\cdot|m_{1 \shortrightarrow 2},s^*)
        \,,\;
        \rho_{B^n}^{(m_{1 \shortrightarrow 2},m_{2 \shortrightarrow 3})}=\mathcal{F}_{M_{1 \shortrightarrow 2}M_{2 \shortrightarrow 3} S\to B^n}(m_{1 \shortrightarrow 2},m_{2 \shortrightarrow 3},s^*)\,,
        \\
        &\rho_{C^n}^{(m_{2 \shortrightarrow 3})}=\mathcal{D}_{M_{2 \shortrightarrow 3}S\to C^n}(m_{2 \shortrightarrow 3},s^*)\,.
\end{align}
which no longer require CR.
\end{proof}

Next, %
we characterize the achievable rates for empirical coordination. We begin with a basic two-node network, and then generalize to   
 a cascade network.  
Based on Theorem~\ref{Theorem: CR theorem} above, introducing 
CR does not affect the achievable rates. Therefore, we will  focus our definitions on  empirical coordination without CR.  

\subsection{Two-node network}
\begin{figure}[tb]
\center
\includegraphics[scale=0.75,trim={5.3cm 0 5.5cm 0}]
{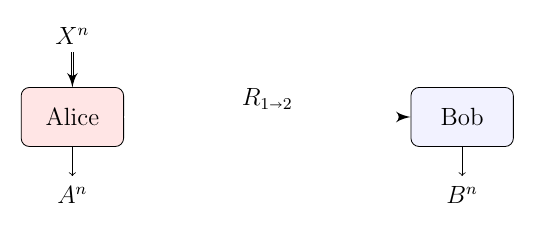} %
\caption{Two-node network with a single quantum link.
}
\label{Figure: Two nodes}
\end{figure}

Consider the two-node network in Figure~\ref{Figure: Two nodes}. 
Alice and Bob would like to simulate a separable state $\omega_{AB}$ on average
using the following coding scheme.
Alice receives classical side information from a memoryless source $p_X$.
She %
encodes $A^n$, and then sends 
 an index $m_{1 \shortrightarrow 2}$, i.e., a classical message to Bob, at a rate $R_{1 \shortrightarrow 2}$. 

Formally, 
a $\left(%
2^{n R_{1 \shortrightarrow 2}},n\right)$
empirical coordination code for  a separable state 
$\omega_{AB}
$ consists
of an input distribution $p_{M_{1 \shortrightarrow 2}|X^n}$ over an index set %
 $ \big[2^{nR_{1 \shortrightarrow 2}}\big]$, 
and two c-q encoding channels, $%
\mathcal{E}_{X\to A}
$ %
and %
$%
 \mathcal{F}_{M_{1 \shortrightarrow 2}\to B^n}
$. %
The %
protocol works as follows.
Alice receives $x^n$, drawn according to $ p_X^n$.
She selects a random index,
\begin{align}
m_{1 \shortrightarrow 2}\sim p_{M_{1 \shortrightarrow 2}|X^n}(\cdot|x^n) %
\end{align}
and sends it through a noiseless link.
Furthermore, she encodes $A^n$ by
\begin{align}
\rho_{A^n}^{(x^n)}=
\bigotimes_{i=1}^n
\mathcal{E}_{X\to A}(x_i) \,.
\end{align}
As Bob receives the message $m_{1 \shortrightarrow 2}$, he prepares the state
\begin{align}
\rho_{B^n}^{(m_{1 \shortrightarrow 2})}=\mathcal{D}_{M_{1 \shortrightarrow 2}\to B^n} (m_{1 \shortrightarrow 2}) \,. 
\end{align}
Hence, the resulting average (joint) state is
\begin{align}
\overline{\rho}_{A B}
=
\frac{1}{n}
\sum_{i=1}^n \sum_{x^n\in \mathcal{X}^n}
\sum_{m_{1 \shortrightarrow 2}\in [2^{n R_{1 \shortrightarrow 2}}]}
p_{X}^n(x^n) p_{M_{1 \shortrightarrow 2}|X^n}(m_{1 \shortrightarrow 2}|x^n)
\rho_{A_i}^{(x^n)}
\otimes  \rho_{B_i}^{
(m_{1 \shortrightarrow 2})}  
\,.
\end{align}

\begin{definition}
A  rate $R_{1 \shortrightarrow 2}\geq 0$
is %
achievable for the empirical coordination of $\omega_{AB}$, if for every $\varepsilon,\alpha>0$ and sufficiently large $n$,
there exists a  $\left(2^{n(R_{1 \shortrightarrow 2}+\alpha)},n\right)$
code that achieves
\begin{align}
\norm{\overline{\rho}_{AB}-\omega_{AB}}_1 \leq \varepsilon  \,.
\end{align}
 \label{Definition:Two_Node_Coordination_Region}  
\end{definition}

\begin{definition}
\label{Definition: Two Node Capacity}
The empirical coordination capacity for the simulation of  a separable state $\omega_{AB}$ over the two-node network is defined as the infimum of achievable rates. We denote the capacity by
$C_{\text{2-node}}(\omega)$.
\end{definition}
    
The optimal rate for empirical coordination is established below.
Consider the extended c-q state,
\begin{align}
\omega_{XAB}=
\sum_{x\in\mathcal{X}}
p_X(x) \ketbra{x}_X\otimes \omega_{AB}^x
\end{align}
where here, $X$ plays the role of a classical register.
Furthermore, let $\mathscr{S}_{\text{2-node}}(\omega)$ be the set of all c-q extensions
\begin{subequations}
\begin{align}
  \sigma_{XYAB}&=
 \sum\limits_{
 (x,y)\in\mathcal{X}\times \mathcal{Y}
 }
 p_{XY}(x,y)  
 \ketbra{x}\otimes\ketbra{y}\otimes \sigma_A^x
  \otimes 
  \sigma_{B}^{y} 
\end{align}
such that %
\begin{align}
 \sigma_{XAB}&=\omega_{XAB} \,.
\end{align}
\end{subequations}

Notice that given a classical pair $(X,Y)=(x,y)$, there is no correlation between $A$ and $B$. 
We also note that if $\omega_{AB}$ is entangled, then
$\mathscr{S}_{\text{2-node}}(\omega)$ is an empty set.

\begin{theorem}
\label{Theorem: Two nodes}
Let $\omega_{AB}$ be a bipartite state in 
$\Delta(\mathcal{H}_A\otimes \mathcal{H}_B)$.
If the set $\mathscr{S}_{\text{2-node}}(\omega)$ is nonempty, then the empirical coordination capacity  for the two-node network  in Figure~\ref{Figure: Two nodes} is given by 
\begin{align}
\label{Equation: Coordinaiton capacity - Two nodes}
C_{\text{2-node}}(\omega)=
\inf_{\sigma\in \mathscr{S}_{\text{2-node}}(\omega)}
I(X;Y)_\sigma 
\,.
\end{align}
Otherwise, if $\mathscr{S}_{\text{2-node}}(\omega) = \emptyset$, then coordination is impossible. 
\end{theorem}

The achievability proof for Theorem~\ref{Theorem: Two nodes} is given in Subsection~\ref{Achievability: Two nodes}, and the converse  in Subsection~\ref{Converse: Two nodes}.

\begin{remark}
\label{Remark: Entanglement}
The set    $\mathscr{S}_{\text{2-node}}(\omega)$ is empty if and only if $\omega_{AB}$ is entangled. As mentioned in Remark~\ref{Remark: Separable},  classical links cannot generate entanglement, hence, coordination is impossible in this case.
\end{remark}

\begin{remark}
\label{Remark: Decomposition}
The characterization involves optimization over a collection of separable states, 
$\mathscr{S}_{\text{2-node}}(\omega)$.
This is a unique feature of the quantum setting. In the classical setting, there is no optimization. 
As will be seen in Examples \ref{Example: Two nodes 1} and \ref{Example: Two nodes 2},
 the performance depends heavily on the chosen decomposition. 
\end{remark}

\begin{remark}
In the special case of orthonormal sets,  $\{
\ket{\sigma_A^x}\}$ and $\{
\ket{\sigma_B^y}\}$,  the coordination capacity satisfies
$%
C_{\text{2-node}}(\omega)=
I(A;B)_\omega 
$. %
This case is essentially classical.
\end{remark}

\begin{remark}
One may always find a decomposition of a separable state into a combination of \emph{pure} states. 
In particular, consider
\begin{align}
\label{Equation: Separable state decomposition}
  \omega_{AB}&=
 \sum\limits_{
 (x,y)\in\mathcal{X}\times \mathcal{Y}
 }
 p_{XY}(x,y) \sigma_A^x \otimes  \sigma_{B}^{y} 
 \,.
\end{align}
By inserting spectral decompositions, 
\begin{align}
\sigma_A^x =\sum_{ v_1\in\mathcal{V}_1 } p_{V_1|X}(v_1|x)
 \ketbra{\psi_A^{x,v_1}} \,,\; %
 \sigma_{B}^{y}=\sum_{ v_2\in\mathcal{V}_2 } p_{V_2|Y}(v_2|y) 
 \ketbra{\phi_B^{y,v_2}}
 \,,
 \end{align}
 we obtain 
\begin{align}
  \omega_{AB}&=
  \sum_{ w_1,w_2 }
 p_{W_1 W_2}(w_1,w_2)
 \ketbra{\psi_A^{w_1}}
 \otimes   
 \ketbra{\phi_B^{w_2}}
\end{align}
where $W_1\equiv (X,V_1)$ and $W_2\equiv (Y,V_2)$. 
If one uses this pure-state decomposition, then the coordination rate would be $R> I(W_1;W_2)_{\sigma}$. Nevertheless, the theorem shows that this can be suboptimal, since 
$I(XV_1;Y V_2)_{\sigma}\geq I(X;Y)_{\sigma} $.
\end{remark}

\begin{remark}
Based on our previous result \cite{NaturPereg_CQ_ITW,natur2025quantum}, strong coordination can be achieved at the same rate if Alice and Bob share sufficient CR before communication begins. Here, however, we assume that CR is not available to Alice and Bob. Yet, they can perform the coordination task at this rate, since the requirement of empirical coordination is less strict.
\end{remark}

\begin{example}%
\label{Example: Two nodes 1}
Let $A$ and $B$ be a qubit pair, i.e., 
$\mathrm{dim}(\mathcal{H}_A)=\mathrm{dim}(\mathcal{H}_B)=2$.
Consider 
the state
\begin{align}
\omega_{AB}=\frac{1}{2}\ketbra{0}\otimes\ketbra{0}
+
\frac{1}{4}
\ketbra{1}\otimes\ketbra{0}
+
\frac{1}{4}
\ketbra{1}\otimes\ketbra{+}
\end{align}
where $\{\ket{0},\ket{1}\}$ and $\{\ket{+},\ket{-}\}$ are the computational basis and conjugate basis, respectively. 
Such decomposition can be associated with a joint distribution $p_{XY}$, where $Y=X$ with probability $1$, an alphabet of size $\abs{\mathcal{X}}=3$,
and 
\begin{align}
p_X =\left( \frac{1}{2}, \frac{1}{4}, \frac{1}{4} \right)\,.
\end{align}
Based on Theorem~\ref{Theorem: Two nodes}, we can achieve
 the rate $R_{1\shortrightarrow 2}=I(X;Y)_{\sigma}=1.5$.
The coordination rate can be significantly improved by using the decomposition below instead, 
\begin{align}
\rho_{AB}=\frac{1}{2}\ketbra{0}\otimes\ketbra{0}
+
\frac{1}{2}
\ketbra{1}\otimes\eta
\end{align}
where $\eta$ is the BB84 state,
\begin{align}
\eta=\frac{1}{2}\ketbra{0}+\frac{1}{2}\ketbra{+}
\,.
\end{align}
This yields the improved rate of $R_{1\shortrightarrow 2}=I(X;Y)_{\sigma}%
=0.3112$.

\end{example}

\begin{example}
\label{Example: Two nodes 2}

Consider the following qubit state, 
\begin{align}
\omega_{AB}=
\frac{1}{2}\ketbra{0}\otimes
\left[(1-p)\ketbra{+}+p\ketbra{-}\right]+
\frac{1}{2}\ketbra{1}\otimes
\left[p\ketbra{+}+(1-p)\ketbra{-}\right]
\end{align}
where the second qubit can be viewed as the output of a phase-flip channel, %
 $p\in (0,1)$.
In this case, we obtain
\begin{align}
I(X;Y)_{\sigma}=1-h(p)
\end{align}
where $h(x)=-(1-x)\log(1-x)-x\log(x)$ is the binary entropy function on $(0,1)$. 
For $p=\frac{1}{2}$, we have a product state
$\omega_{AB}=\frac{\identity}{2}\otimes \frac{\identity}{2}$.
Hence, communication is not necessary and the coordination capacity is $C_{\text{2-node}}(\omega)=0$.

\end{example}

\subsection{Cascade network}
\begin{figure}[tb]
\center
\includegraphics[scale=0.75,trim={5.3cm 0 5.5cm 0}]
{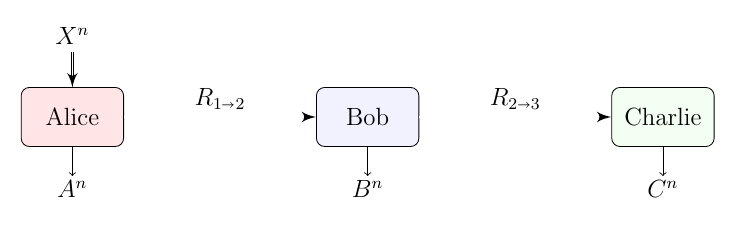} %
\caption{Cascade network with classical links and without common randomness.
}
\label{Figure: Cascade network}
\end{figure}

Consider the cascade network (see Figure~\ref{Figure: Cascade network}). 
Alice, Bob, and Charlie wish to simulate a separable state $\omega_{ABC}$
using the following scheme.
Alice receives classical side information from a memoryless source $p_X$.
She %
encodes $A^n$, and she sends 
 an index $m_{1 \shortrightarrow 2}$, i.e., a classical message to Bob, at a rate $R_{1 \shortrightarrow 2}$. Then Bob uses the message $m_{1 \shortrightarrow 2}$ to encode his systems $B^n$, and sends a message $m_{2 \shortrightarrow 3}$ to Charlie who uses it to encode his systems $C^n$.

Formally, 
a $\left(
2^{n R_{1 \shortrightarrow 2}},2^{n R_{2 \shortrightarrow 3}},n\right)$
empirical coordination code for the 
simulation of a separable state 
$\omega_{ABC}$ in the cascade network consists
two input distributions $p_{M_{1 \shortrightarrow 2}|X^n}$ and $p_{M_{2 \shortrightarrow 3}|X^nM_{1 \shortrightarrow 2}}$  over index sets %
 $ \big[2^{nR_{1 \shortrightarrow 2}}\big] $ and $\big[2^{nR_{2 \shortrightarrow 3}}\big]$, 
and three encoding c-q channels $%
\mathcal{E}_{X\to A}%
\Delta(\mathcal{H}_A)
$, $
\mathcal{F}_{M_{1 \shortrightarrow 2}\to B^n} %
$, and $\mathcal{D}_{M_{2 \shortrightarrow 3}\to C^n} %
$.  %
The %
protocol works as follows.

Alice selects a random index,
\begin{align}
m_{1 \shortrightarrow 2}\sim p_{M_{1 \shortrightarrow 2}} %
\end{align}
and sends it through a noiseless link.
Furthermore, she encodes $A^n$ by
\begin{align}
\rho_{A^n}^{(x^n)}=\bigotimes_{i=1}^{n}
\mathcal{E}_{X\to A}(x_i) \,.
\end{align}
As Bob receives the message $m_{1 \shortrightarrow 2}$, he generates  $m_{2 \shortrightarrow 3}$ according to 
$p_{M_{2 \shortrightarrow 3}|X^nM_{1 \shortrightarrow 2}}(\cdot|x^n,m_{1\shortrightarrow 2})$,
sends $m_{2 \shortrightarrow 3}$ to Charlie,
and prepares the state
\begin{align}
\rho_{B^n}^{(m_{1 \shortrightarrow 2})} =\mathcal{F}_{M_{1 \shortrightarrow 2}\to B^n} (m_{1 \shortrightarrow 2}) \,. 
\end{align}
Having received the classical message $m_{2 \shortrightarrow 3}$,  Charlie applies his  c-q encoding map and prepares %
\begin{align}
\rho_{C^n}^{(m_{2 \shortrightarrow 3})}=
\mathcal{D}_{M_{2 \shortrightarrow 3}\to C^n}(m_{2 \shortrightarrow 3}) \,.
\end{align}
Hence, the resulting average (joint) state is
\begin{multline}%
\overline{\rho}_{A B C}
=
\sum_{x^n\in \mathcal{X}^n} p_{X}^n(x^n) 
\sum_{m_{1 \shortrightarrow 2}\in [2^{n R_{1 \shortrightarrow 2}}]} \sum_{m_{2 \shortrightarrow 3}\in [2^{n R_{2 \shortrightarrow 3}}]}
p_{M_{1 \shortrightarrow 2}|X^n}(m_{1 \shortrightarrow 2}|x^n) p_{M_{2 \shortrightarrow 3}|M_{1 \shortrightarrow 2}X^n}(m_{2 \shortrightarrow 3}|m_{1 \shortrightarrow 2},x^n)
\\
\cdot
\frac{1}{n}
\sum_{i=1}^n
\rho_{A_i}^{(x^n)}
\otimes  \rho_{B_i}^{
(m_{1 \shortrightarrow 2})} \otimes \rho_{C_i}^{(m_{2 \shortrightarrow 3})}  \,.
\end{multline}%

\begin{definition}
A  rate pair $(R_{1 \shortrightarrow 2},R_{2 \shortrightarrow 3})$
is %
achievable for the empirical coordination of  $\omega_{ABC}$ %
if for every $\varepsilon,\delta>0$ and a sufficiently large $n$,
there exists a  $\left(2^{n(R_{1 \shortrightarrow 2}+\delta)},2^{n(R_{2 \shortrightarrow 3}+\delta)},n\right)$
code that achieves
\begin{align}
\norm{\overline{\rho}_{ABC}-\omega_{ABC}}_1 \leq \varepsilon  \,.
\end{align}
\end{definition}

\begin{definition}
The empirical coordination capacity region  for
the simulation of a separable state $\omega_{ABC}$ over
the cascade network is defined as the closure of all the achievable rate pairs $(R_{1,2},R_{2,3})$.

We denote the capacity region by $\mathcal{C}_{\text{Cascade}}(\omega)$.
\end{definition}

The main result for the cascade network is established below. Consider the extended c-q state,
\begin{align}
\omega_{XABC}=
\sum_{x\in\mathcal{X}}
p_X(x) \ketbra{x}_X\otimes \omega_{ABC}^x \,.
\end{align}
Furthermore, let $\mathscr{S}_{\text{Cascade}}(\omega)$ be the set of all c-q extensions
\begin{subequations}
\begin{align}
  \sigma_{XYZABC}&=
 \sum\limits_{
 (x,y,z)\in\mathcal{X}\times \mathcal{Y}\times \mathcal{Z}
 }
 p_{XYZ}(x,y,z)  
 \ketbra{x}\otimes\ketbra{y}\otimes\ketbra{z}\otimes \sigma_A^x
  \otimes 
  \sigma_{B}^{y}
  \otimes 
  \sigma_{C}^{z}
\end{align}
such that %
\begin{align}
 \sigma_{XABC}&=\omega_{XABC} \,.
\end{align}
\end{subequations}

As before, coordination with classical links is limited to separable states (see Remarks~\ref{Remark: Separable} and \ref{Remark: Entanglement}).
\begin{theorem}
\label{Theorem: Cascade}
Let $\omega_{ABC}$ be a tripartite state in 
$\Delta(\mathcal{H}_A\otimes \mathcal{H}_B\otimes \mathcal{H}_C)$.
    If the set $\mathscr{S}_{\text{Cascade}}(\omega)$ is nonempty, then the empirical coordination capacity  region for the cascade network  in Figure~\ref{Figure: Cascade network} is 
\begin{align}
\mathcal{C}_{\text{Cascade}}(\omega)=
\bigcup_{ 
\mathscr{S}_{\text{Cascade}}(\omega)}
\left\{ 
\begin{array}{rrl}
\left(R_{1 \shortrightarrow 2}, R_{2 \shortrightarrow 3}\right):
&
R_{1 \shortrightarrow 2}        &\geq I(X;YZ)_\sigma \,,\\
&R_{2 \shortrightarrow 3}  &\geq I(X;Z)_\sigma
\end{array}
\right\} \,.
\end{align}
Otherwise, if $\mathscr{S}_{\text{Cascade}}(\omega) = \emptyset$, then coordination is impossible.
\end{theorem}
The achievability  proof for Theorem~\ref{Theorem: Cascade} is provided in Subsection~\ref{Achievability: Cascade}, and the converse part  in Subsection~\ref{Converse: Cascade}.

\begin{remark}
The cascade model has a Markov structure in the sense that given the message $m_{2\to 3}$ from Bob, Charlie's state $\rho_{C^n}^{m_{2\to 3}}$ has no correlation with Alice.
Nevertheless, the correlation that Alice, Bob, and Charlie simulate does not satisfy a Markov chain property. In particular, the auxiliary random variables $X$, $Y$, and $Z$ may follow a general Bayesian rule, and
do not necessarily form a Markov chain.  
\end{remark}

\subsection{Isolated node}

\begin{figure}[tb]
\center
\includegraphics[scale=0.75,trim={5.3cm 0 5.5cm 0}]
{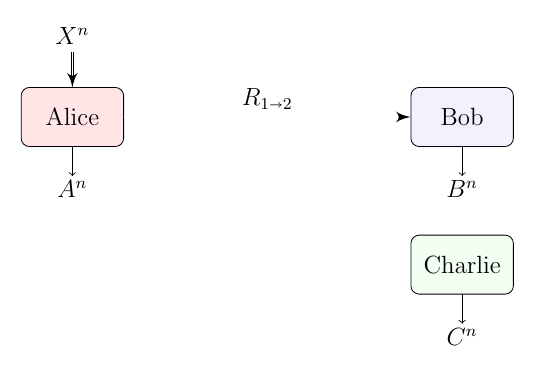} %
\caption{Isolated node network with a single classical link.
}
\label{Figure: Isolated node network}
\end{figure}

Consider the isolated node network in Figure~\ref{Figure: Isolated node network}.
This a special case of a cascade network with $R_{2\shortrightarrow 3}=0$.
The coordination capacity $C_{\text{Isolated}}(\omega)$
is defined similarly as in Definition~\ref{Definition: Two Node Capacity}, and can be established as a consequence of 
Theorem~\ref{Theorem: Cascade}.
Consider the extended c-q state,
\begin{align}
\omega_{XABC}=
\sum_{x\in\mathcal{X}}
p_X(x) \ketbra{x}_X\otimes \omega_{ABC}^x \,.
\end{align}
Let $\mathscr{S}_{\text{Isolated}}(\omega)$ be the set of all c-q extensions $\sigma_{XYZABC}$ of the form
\begin{subequations}
\begin{align}
  \sigma_{XYZABC}&=
 \sum\limits_{
 (x,y,z)\in\mathcal{X}\times \mathcal{Y}\times \mathcal{Z}}
 p_{XYZ}(x,y,z)  
 \ketbra{x}\otimes\ketbra{y}\otimes\ketbra{z}\otimes \sigma_A^x
  \otimes 
  \sigma_{B}^{y}
  \otimes 
  \sigma_{C}^{z}
\end{align}
such that %
\begin{align}
 \sigma_{XABC}&=\omega_{XABC}
 \intertext{and}
 \sigma_{AC}&=\sigma_A\otimes \sigma_C
 \,.
\end{align}
\end{subequations}

\begin{corollary}
\label{Corrolary: Isolated node}
Let $\omega_{ABC}$ be as in Theorem~\ref{Theorem: Cascade}.
If the set $\mathscr{S}_{\text{Isolated}}(\omega)$ is nonempty, then the empirical coordination capacity  for the isolated node network  in Figure~\ref{Figure: Isolated node network} is given by 
\begin{align}
\label{Equation: Coordinaiton capacity - Cascade}
C_{\text{Isolated}}(\omega)=
\inf_{\sigma\in \mathscr{S}_{\text{Isolated}}(\omega)}
I(X;Y|Z)_\sigma 
\end{align} 
Otherwise, if $\mathscr{S}_{\text{Isolated}}(\omega) = \emptyset$, then coordination is impossible. 
\end{corollary}
In this case, coordination is only possible for a separable state $\omega_{ABC}$ such that $\omega_{AC}=\omega_A\otimes \omega_C$.

\begin{remark}
    Notice that $B$ and $C$ can still be correlated.
    Given unlimited CR, it is clear that we may generate such a correlation.
    Even in the extreme case of no communication, we can
    generate $Y^n$ from a memoryless source, treat $Y^n$ as the CR element, and let $Z^n=Y^n$
    (see discussion in \cite[Sec. III-B]{cuff2010coordination}). We have seen that CR does not affect the coordination capacity, and thus, the same rates can be achieved without CR. Further intuition is given in the discussion in Subsection~\ref{Section: CR}.
\end{remark}
\begin{example}
\label{Example: Cascade}

Consider the following qubit state, 
\begin{align}
\omega_{ABC}&=
(1-\alpha)\ketbra{0}\otimes
\big[(1-p)\ketbra{+}\otimes \ketbra{+}
+p\ketbra{-}\otimes \ketbra{-}\big]
\nonumber 
\\
&
+
\alpha\ketbra{1}\otimes
\big[(1-p)\ketbra{+}\otimes \ketbra{-}
+p\ketbra{-}\otimes \ketbra{+}\big]
\end{align}
with  %
 $\alpha,p\in (0,1)$.
In this case, 
$%
I(X;Y|Z)_{\sigma}=H(X)=h(\alpha)
$. %

\end{example}

\section{Main Results - Quantum Links}
\label{Section: QL Results}
\subsection{Two-node network}
\label{Subsection:Two-Nodes}
\begin{figure}[tb]
\center
\includegraphics[scale=0.75,trim={5.3cm 0 5.5cm 0}]
{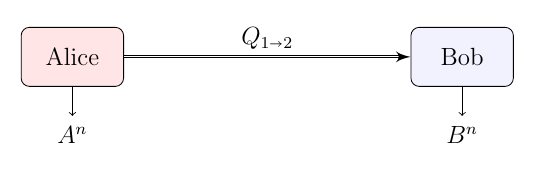} %
\caption{Two-node network with a quantum link.
}
\label{Figure: Two-node network - Empirical - Quantum links}
\end{figure} 
Consider the two-node network with quantum links presented in Figure~\ref{Figure: Two-node network - Empirical - Quantum links}. 
Alice and Bob would like to simulate a  state $\omega_{AB}\in \Delta(\mathcal{H}_{A}\otimes\mathcal{H}_{B})$. Let $\ket{\omega_{RAB}}$ be a purification of $\omega_{AB}$, where $R$ can be viewed as Alice's reference system. 
To achieve empirical coordination with respect to $\omega_{AB}$, they apply the scheme below.
Alice 
encodes her systems $A^n$, and a quantum description $M_{1 \rightarrow 2}$ to Bob, at a rate $Q_{1 \rightarrow 2}$. 

Formally, 
a $\left(%
2^{n Q_{1 \rightarrow 2}},n\right)$
empirical coordination code for the two-node network with quantum links in Figure~\ref{Figure: Two-node network - Empirical - Quantum links}
consists
of a Hilbert space $M_{1\rightarrow 2}$ of dimension
\begin{align}
    \dim(\mathcal{H}_{M_{1\rightarrow 2}})=2^{nQ_{1\rightarrow 2}}
\end{align}
and two encoding channels 
\begin{align}
\mathcal{E}_{\bar{A}^n\bar{B}^n\to A^n M_{1\rightarrow2}} \,:\;& \Delta(\mathcal{H}_{A}^{\otimes n}\otimes\mathcal{H}_B^{\otimes n})\to\Delta(\mathcal{H}_{A}^{\otimes n}\otimes\mathcal{H}_{M_{1\rightarrow2}})\,,
\\
\mathcal{F}_{M_{1 \rightarrow 2}\to B^n} \,:\;& \Delta(\mathcal{H}_{M_{1 \rightarrow 2}})\to\Delta(\mathcal{H}_B^{\otimes n})
\end{align}
The %
protocol works as follows.
First, Alice prepares the state $\omega_{R\breve{A}}^{\otimes n}$ locally, where $\breve{A}$ is her ancilla.
She  applies her encoding map $\mathcal{E}_{\breve{A}^n\to A^n M_{1\rightarrow2}}$ producing the following state,
\begin{align}
   \tau_{R^{n}A^{n}M_{1\rightarrow 2}}=
(\mathrm{id}_{R^n}\otimes\mathcal{E}_{\breve{A}^n \to A^{n}M_{1\rightarrow 2}})(\omega_{R\breve{A}}^{\otimes n} )
\,,. 
\end{align}
Alice then sends the ``quantum message" $M_{1\rightarrow2}$ to Bob
 through a noiseless quantum link.

As Bob receives  $M_{1 \rightarrow 2}$, he applies his decoding channel $\mathcal{F}_{M_{1 \rightarrow 2}\to B^n}$. The resulting state is 
\begin{align}\rho_{R^nA^nB^n}=\left(\mathrm{id_{R^nA^n}}\otimes\mathcal{F}_{M_{1 \rightarrow 2}\to B^n} \right)(\tau_{R^nA^nM_{1\rightarrow 2}}) \,. 
\end{align}

\begin{definition}

A  rate $Q_{1 \rightarrow 2}\geq 0$
is %
achievable for the empirical coordination of $\omega_{RAB}$, if for every $\varepsilon,\alpha>0$ and sufficiently large $n$,
there exists a  $\left(2^{n(Q_{1 \rightarrow 2}+\alpha)},n\right)$
code that achieves
\begin{align}
\norm{\frac{1}{n}\sum_{i=1}^n {\rho}_{R_iA_iB_i}-\omega_{RAB}}_1 \leq \varepsilon  \,.
\end{align}
\end{definition}

\begin{definition}
The empirical coordination capacity for the simulation of  a state $\omega_{AB}$ over the two-node network with quantum links is defined as the infimum of achievable rates. We denote the capacity by
$\mathcal{Q}_{\text{2-node}}(\omega)$.
\end{definition}
    
The optimal rate for empirical coordination is established below.

\begin{theorem}
\label{Theorem: Two-node - Quantum links}
Let $\omega_{AB}$ be a bipartite state in 
$\Delta(\mathcal{H}_A\otimes \mathcal{H}_B)$ with a purification $\ket{\omega_{RAB}}$.
The empirical coordination capacity  for the two-node network with quantum links in Figure~\ref{Figure: Two-node network - Empirical - Quantum links} is given by 
\begin{align}
\mathcal{Q}_{\text{2-node}}(\omega)=H(B)_\omega
\,.
\end{align}
\end{theorem}

The  proof for Theorem~\ref{Theorem: Two-node - Quantum links} is given in Subsection~\ref{Converse: Two nodes - Empirical - Quantum links}.
The achievability part follows from the strong  coordination result~\cite{schumacher1995quantum}.
The converse proof is the challenging part of the analysis. %

\begin{remark}
\label{Remark: Schumacher}
Based on the result in Theorem~\ref{Theorem: Two-node - Quantum links}, we observe that
the Schumacher compression protocol is optimal. That is, Alice can simply prepare $\omega_{AB}^{\otimes n}$ locally, compress $B^n$ into a quantum representation $M_{1\to 2}$ and send to Bob at a rate of $Q_{1\to 2}= H(B)_\omega+\delta$,  and then Bob decodes as usual. %
\end{remark}

\subsection{Broadcast network}
\begin{figure}[tb]
\center
\includegraphics[scale=0.75,trim={5.3cm 0 5.5cm 0}]
{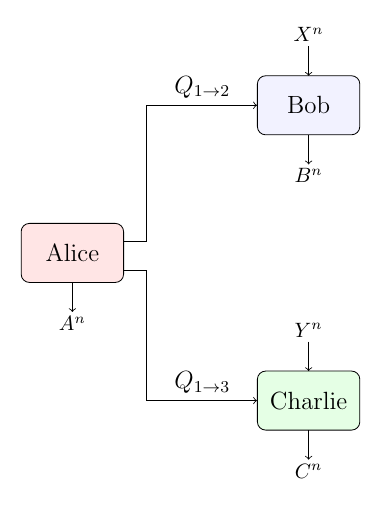} %
\caption{Broadcast network with quantum links.
}
\label{Figure: Broadcast network - Empirical - Quantum links}
\end{figure} 
Consider the broadcast network with quantum links presented in Figure~\ref{Figure: Broadcast network - Empirical - Quantum links}. 
Consider a c-c-q-q-q state, %
\begin{align}
\omega_{XYABC}=\sum_{x\in\mathcal{X}} \sum_{y\in\mathcal{Y}}
p_{XY}(x,y)\ketbra{x,y}_{X,Y}
\otimes \ketbra{\omega^{(x,y)}_{ABC}}
\label{Equation:Broadcast_omega_XYABC}
\end{align}
where $X,Y$ are classical registers that store the values $x,y$, respectively, drawn from a joint memoryless source $p_{XY}$. 

Alice, Bob, and Charlie would like to simulate the state $\omega_{XYABC}$ on average. Before communication takes place,
the classical sequences $X^n$ and $Y^n$ are drawn from a common source $p_{XY}^{\otimes n}$.
The sequence $X^n$ is given to Bob, while $Y^n$ is given to Charlie as illustrated in Figure~\ref{Figure: Broadcast network - Empirical - Quantum links}. 
To achieve empirical coordination with respect to $\omega_{XYABC}$, they apply the following scheme.  
Initially, Alice prepares the state of her output $A^{n}$, along with two quantum descriptions, $M_{1\rightarrow 2}$ and $M_{1 \rightarrow 3}$. She then transmits   $M_{1 \rightarrow 2}$ and $M_{1 \rightarrow 3}$, to Bob and Charlie, respectively,  at limited qubit transmission rates, $Q_{1\rightarrow 2}$ and $Q_{1\rightarrow 3}$.
As Bob receives the quantum description $M_{1\rightarrow 2}$, he uses it together with the classical sequence  $X^n$ to   encode the output $B^{n}$, i.e., apply an encoding map to configure the output state. Similarly, Charlie receives $M_{1\rightarrow 3}$ and  $Y^n$, and encodes his output $C^{n}$. 

Formally, 
a $\left(%
2^{n Q_{1 \rightarrow 2}},2^{n Q_{1 \rightarrow 3}},n\right)$
empirical coordination code for the two-node network with quantum links in Figure~\ref{Figure: Two-node network - Empirical - Quantum links}
consists of 
two Hilbert spaces, $\mathcal{H}_{M_{1 \rightarrow 2}}$ and $\mathcal{H}_{M_{1\rightarrow 3}}$, of dimensions
\begin{align} 
\mathrm{dim}(\mathcal{H}_{M_{1 \rightarrow 2}})=2^{nQ_{1\rightarrow 2}},\;
\mathrm{dim}(\mathcal{H}_{M_{1 \rightarrow 3}})=2^{nQ_{1\rightarrow 3}}
\end{align} 
and three encoding maps,
\begin{align}
\mathcal{E}_{A^{n}\to A^{n}M_{1\rightarrow 2}M_{1 \rightarrow 3}} &: \Delta(\mathcal{H}_{A}^{\otimes n}) \to \Delta(\mathcal{H}_{A}^{\otimes n} \otimes \mathcal{H}_{M_{1 \rightarrow 2}} \otimes \mathcal{H}_{M_{1 \rightarrow 3}}),
\\
\mathcal{F}_{X^n M_{1\rightarrow 2}\to B^{n}} &: \mathcal{X}^n\otimes  \Delta(\mathcal{H}_{  M_{1\rightarrow 2}}) \to \Delta(\mathcal{H}_{ B}^{\otimes n}),
\intertext{and}
\mathcal{D}_{Y^nM_{1 \rightarrow 3}\to  C^{n}} &: \mathcal{Y}^n\otimes \Delta(\mathcal{H}_{ M_{1\rightarrow 3}}) \to \Delta(\mathcal{H}_{C}^{\otimes n}).
\end{align}%
The %
protocol works as follows.
Alice first prepares the state 
$\omega_{\bar{A}}^{\otimes n}$, where $\bar{A}^n$ is her ancilla.
She then applies her encoding map %
and prepares %
\begin{align} 
\rho_{A^{n}M_{1 \rightarrow 2}M_{1\rightarrow 3}}=
 \mathcal{E}_{A^{n}\to A^{n}M_{1 \rightarrow 2}M_{1 \rightarrow 3}}(\omega_{A}^{\otimes n}) \,. 
\end{align}
She sends  $M_{1 \rightarrow 2}$ and $M_{1 \rightarrow 3}$ to Bob and Charlie, respectively. Once Bob receives $M_{1 \rightarrow 2}$ and the classical assistance, $X^n$, he applies his encoding map 
$\mathcal{F}_{X^n M_{1 \rightarrow 2}\to B^{n}}$. 
Similarly,  Charlie receives $M_{1 \rightarrow 3}$ and $Y^n$, and applies $\mathcal{D}_{Y^nM_{1 \rightarrow 3}\to  C^{n}}$. Their encoding operations result in the following extended state:
\begin{multline}
\widehat{\rho}_{X^n Y^n A^{n}B^{n}C^{n}}
=
\sum_{x^n\in\mathcal{X}^n} \sum_{y^n\in\mathcal{Y}^n}
 p_{XY}^{\otimes n}(x^n,y^n) \ketbra{x^n,y^n}_{X^n Y^n}\otimes
 \\
 (\mathrm{id}_{A^{n}}\otimes\mathcal{F}_{X^{n}M_{1 \rightarrow 2}\to B^{n}}\otimes\mathcal{D}_{Y^{n}M_{1 \rightarrow 3}\to C^{n}})
 \left(
 \ketbra{x^n,y^n}_{\bar{X}^n \bar{Y}^n} \otimes\rho_{A^{n}M_{1 \rightarrow 2}M_{1 \rightarrow 3}}^{(1)}
 \right)
\,,
\end{multline}
where $\bar{X}^n \bar{Y}^n$ are classical registers that store a copy of the (classical) sequences $X^n Y^n$, respectively.

\begin{definition}
A rate pair $(Q_{1 \rightarrow 2},Q_{1 \rightarrow 3})$ is achievable for empirical coordination with quantum links, if for every $\varepsilon,\delta>0$ and a 
sufficiently large $n$, there exists a $(2^{n(Q_{1 \rightarrow 2}+\delta)},2^{n(Q_{1 \rightarrow 3}+\delta)},n)$ coordination code satisfying 
\begin{align}
\norm{\frac{1}{n}\sum_{i=1}^n \widehat{\rho}_{X_iY_iA_iB_iC_i}-\omega_{XYABC}}_1 \leq \varepsilon  \,.
\end{align}
\end{definition}

\begin{definition}
The empirical coordination capacity for the simulation of  a state $\omega_{XYABC}$ over the broadcast network with quantum links is defined as the closure of all achievable rate pairs. We denote the capacity by
$C_{\text{Broadcast}}(\omega)$.
\end{definition}
    
The optimal rate for empirical coordination is established below.

\begin{remark}
 \label{Remark:Broadcast_no_correlation}
Since Alice has no knowledge of $X^n$ nor $Y^n$,
 coordination can only be achieved for states 
$\omega_{XYABC}$ such that  
$A$ and $XY$ are uncorrelated, i.e., %
\begin{subequations}
\label{Equation:Broadcast_assumptions}
\begin{align}
\omega_{XYA}&=\omega_{XY}\otimes \omega_A \,
\label{Equation:Broadcast_XYA}
\end{align}
Furthermore, Bob does not have access to $Y^n$, whereas Charlie does not know $X^n$. Therefore, the reduced states  $\omega_{AB}^{(x,y)}$ and $\omega_{AC}^{(x,y)}$ must satisfy the following non-signaling properties: %
\begin{align}
\omega_{AB}^{(x,y)}&=\omega_{B}^{(x,y')} \;\text{for all $x\in\mathcal{X}$ and $y,y'\in\mathcal{Y}$} \,,
\label{Equation:Bob_Non_Signaling}\\
\omega_{AC}^{(x,y)}&=\omega_{C}^{(x',y)} \;\text{for all $x,x'\in\mathcal{Y}$ and $y\in\mathcal{Y}$} \,.
\label{Equation:Charlie_Non_Signaling}
\end{align}
\label{Equation:Broadcast_Non_Signaling}
\end{subequations}
 \end{remark}

\begin{theorem}
\normalfont
\label{Theorem: Broadcast - Quantum links}
Let $\omega_{XYABC}$ be as in \eqref{Equation:Broadcast_omega_XYABC}, such that \eqref{Equation:Broadcast_assumptions} holds.
The coordination capacity region for the broadcast network  in Figure~\ref{Figure: Broadcast network - Empirical - Quantum links} is given by the set
\begin{align}
\mathcal{Q}_{\text{Broadcast}}(\omega)=
\left\{ 
\begin{array}{rrl}
\left(Q_{1 \rightarrow 2}, Q_{1 \rightarrow 3}\right)\in\mathbb{R}^{2}:
&
Q_{1\rightarrow 2}        &\geq H(B|X)_\omega \,,\\
&Q_{1\rightarrow 2}  &\geq H(C|Y)_\omega
\end{array}
\right\} \,.
\end{align}
\end{theorem}
The achievability proof for 
Theorem~\ref{Theorem: Broadcast - Quantum links} follows from the result of the broadcast network obtained in the strong coordination framework in \cite{NaturPereg_EC_ITW,natur2025quantum}. The converse part is provided in Section~\ref{Section: Analysis - Broadcast - Empirical - Quantum links}.

\section{Nonlocal Games}
\label{Section:Nonlocal_games}
 In this section, we discuss the connection between empirical  coordination,  nonlocal correlations, and refereed games. %
 We begin with a brief review of refereed games in Subsection~\ref{Subsection: Refereed games}. We explain how coordination is useful in the game realization in Subsection~\ref{Subsection: Coordination in games}. We demonstrate the implications for %
 the CHSH game in Subsection~\ref{Subsection: CHSH}. %
 We follow a similar route as in our discussion for strong coordination in \cite[Sec. V]{natur2025quantum}. Here, however, we focus on empirical coordination, which is relevant for the game
realization.
 
\subsection{Nonlocal Correlations}
\label{Subsection: Refereed games}

\begin{figure}[t]
    \centering
    \hspace{0.5cm}%
    \begin{minipage}[b]{0.3\linewidth} %
        \centering
        \includegraphics[scale=0.7,trim={5.3cm 0 5.5cm 0}]
        {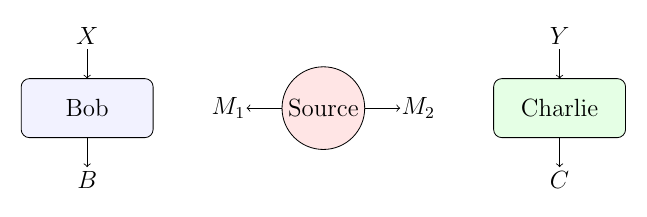} %
        \begin{center}
        \footnotesize{(a) Bell experiment setup.}%
        \end{center}
    \end{minipage}
    \hspace{3.5cm}%
    \begin{minipage}[b]{0.3\linewidth} %
        \centering
         \includegraphics[scale=0.7,trim={5.3cm -1cm 5.5cm 0}]
        {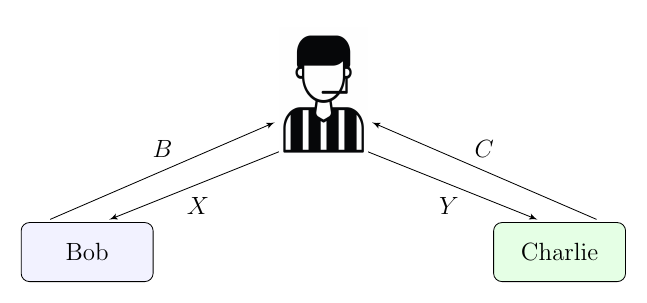} %
        \begin{center}
        \footnotesize{(b) Refereed game setting.}%
        \end{center}
    \end{minipage}
    \caption{Bell experiments and refereed games. Figure~(a) describes a Bell experiment setup, consisting of a source, and two observers. The source distributes physical systems $M_{1\rightarrow 2}$ and $M_{1\rightarrow 3}$ to Bob and Charlie respectively. Bob and Charlie choose to perform measurements $X$ and $Y$ each on his system, yielding classical results $B$ and $C$, respectively. Figure~(b) describes a refereed game where a referee plays against both Bob and Charlie. The referee sends his questions $X$ and $Y$ to Bob and Charlie, they respond with answers $B$ and $C$ respectively. The round is won if the realization of the tuple $(X,Y,B,C)=(x,y,b,c)$ satisfies a specific condition set by the game rules.}
     \label{Figure: Bell experiments and refereed games}
\end{figure}

Nonlocal games are closely related to Bell experiments and quantum correlations \cite{brunner2014bell}.
In the typical setting for a Bell experiment, %
a source distributes two physical systems, $M_{1\rightarrow 2}$ and $M_{1\rightarrow 3}$, to two distant users. See Figure~\ref{Figure: Bell experiments and refereed games}~(a). 
Here, we refer to the users as Bob ($B$) and Charlie ($C$). Upon receiving $M_{1\rightarrow 2}$ and $M_{1\rightarrow 3}$, each chooses to perform a measurement from a particular set of measurements. Denote the measurements  Bob and Charlie chose by $X$ and  $Y$, respectively. The measurements yield the respective outcomes, $B$ and $C$. Notice that $B$ and $C$ are classical in this setting. 

By the nature of quantum measurements \cite{wilde2017quantum}, the outcomes $B$ and $C$ may change from one run of the experiment to another, even when the same measurements $X$ and $Y$ are taken. The outcomes are governed by a conditional probability mass function $P_{BC|XY}(b,c|x,y)$, and can be estimated by
the \emph{empirical average} over a sufficient number of rounds of running the experiment (see Subsection~\ref{Subsection: Justification}). %
The function $P_{BC|XY}$ is also called a behavior, or, \emph{a correlation}.
In general, the correlation cannot necessarily be separated as $%
P_{B|X}\times %
P_{C|Y}%
$, even when the observers are remote. %
This does not necessarily imply a direct influence of one system on the other.

The notion of locality refers to a situation where past factors can be encapsulated in some random variable $U$, also referred to as a hidden variable \cite{einstein1935can,bell1966problem}, such that when taking it into account, the correlation between the outcomes is broken, i.e., %
\begin{align}
    P_{BC|XY}(b,c|x,y)=\int_{\text{supp}(U)}  {p_{U}(u)P_{B|XU}(b|x,u)P_{C|YU}(c|y,u) \,\text{d}u}\,.
    \label{Equation: locality}
\end{align}

The %
predictions of the quantum theory for certain settings
involving quantum entanglement do not follow the locality condition in \eqref{Equation: locality}.
Suppose that Bob and Charlie share a bipartite state $\rho_{M_{1\rightarrow 2} M_{1\rightarrow 3}}$.
As they perform local measurements $\{F_{b|x}\,,\; b\in\mathcal{B}\}$ and 
$\{D_{c|y}\,,\; c\in\mathcal{C}\}$, they generate the following correlation: 
\begin{align}
    P_{BC|XY}(b,c|x,y)=\Tr\left[\left(F_{b|x}\otimes D_{c|y}\right)\rho_{M_{1\rightarrow 2}M_{1\rightarrow 3}}\right]\,.
\label{Equation:Enatnaglement_Correlation}
\end{align}
One of the simplest experiments demonstrating nonlocal behavior is the CHSH  setting, named after Clauser, Horne, Shimony, and Holt \cite{clauser1969proposed}. Consider the Bell experiment setting shown in Figure~\ref{Figure: Bell experiments and refereed games}~(a), where the observers Bob and Charlie can only perform one of two measurements, $X,Y \in \{0,1\}$. The outcomes are limited to two values as well $B,C\in\{\pm 1\}$. Consider 
\begin{align}
    S=\langle B_0 C_0\rangle + \langle B_0 C_1\rangle +\langle B_1 C_0\rangle -\langle B_1 C_1\rangle
    \label{Equation: S quantity}
\end{align}
where $\langle B_x C_y\rangle$ are the corresponding expectation values,
$%
    \langle B_x C_y\rangle=\sum_{b,c\in \{\pm 1\}} bc \cdot P_{BC|XY}(b,c|x,y)
$, for $(x,y)\in\mathcal{X}\times\mathcal{Y}$. %

If the correlation $P_{BC|XY}$ satisfies the locality condition  in \eqref{Equation: locality}, then   $S\leq2$ must hold \cite{bell1964einstein}. However, in the quantum case,  this inequality may be violated. %
Suppose Bob and Charlie are  each provided with a qubit from an EPR pair $\ket{\Phi_{M_{1\rightarrow 2}M_{1\rightarrow 3}}}=\frac{1}{\sqrt{2}}\left(\ket{00}+\ket{11}\right)$. Denote the Pauli operators by $\left(\Sigma_1,\Sigma_2,\Sigma_3\right)$. Bob and Charlie choose their measurements depending on the values of $X$ and $Y$, respectively. %
If $X=0$, Bob measures the $\Sigma_3$ observable. Otherwise, if $X=1$, he measures the $\Sigma_1$ observable. As for Charlie, if $Y=0$, he measures the observable $\frac{-\Sigma_3-\Sigma_1}{\sqrt{2}}$, and if $Y=1$, he measures $\frac{\Sigma_3-\Sigma_1}{\sqrt{2}}$.
This %
yields $S=2\sqrt{2}>2$ (see \eqref{Equation: S quantity}), demonstrating the nonlocal nature of quantum entanglement. Based on 
this violation, quantum correlations cannot be explained using the  theory of classical hidden variables \cite{bell1964einstein}.   

\subsection{Refereed games}
Refereed games can be viewed as another representation of the Bell setting.
Specifically, 
consider the refereed game in Figure~\ref{Figure: Bell experiments and refereed games}~(b). %
The referee provides two  questions $X\in \mathcal{X}$ and $Y\in \mathcal{Y}$, according to some probability distribution $p_{XY}$.
He sends $X$ to the first player (Bob), and $Y$ to the second (Charlie).
Upon receiving their question,  Bob and Charlie respond with classical answers $B\in \mathcal{B}$ and $C\in \mathcal{C}$, respectively. We note that the alphabets $\mathcal{X}$, $\mathcal{Y}$, $\mathcal{B}$, and $\mathcal{C}$ are assumed to be finite. 
The referee %
decides that the game is won if the realization of the tuple $(X,Y,B,C)$ satisfies a specific condition $\mathscr{W}$, set by the rules of the game. This condition is represented by an indicator function,
\begin{align}
    V(x,y,b,c) =
    \begin{cases} 
      1 &  \text{If $(x,y,b,c)$ satisfy $\mathscr{W}$}\,, \\
      0 & \text{otherwise }\,.
\end{cases}
    \label{Equation: Indicator}
\end{align}

We refer to the procedure above %
as a single-shot game. %
 We now discuss the game implementation and rules
 (see also \cite[Sec. V-B]{natur2025quantum}).

\subsubsection{Resources}
As in the Bell setting,  a source distributes correlated physical systems before the procedure begins (see Figure~\ref{Figure: Bell experiments and refereed games}~(a)).
Here, we refer to the source of the correlation resources as \emph{Alice}. 

$\,$

\subsubsection{Strategy}
Before the game starts, i.e., before the referee has chosen his questions, Alice, Bob, and Charlie  meet and agree on a game strategy and the required correlation resources. %
The optimal game strategy and the required correlations for the strategy implementation depend on the game rules. %

$\,$

\subsubsection{No signaling}
During  the course of the game, Bob and Charlie cannot communicate with each other. 
They can, however, exploit the correlation resources in order to coordinate their answers  through quantum measurements.

$\,$

We can also give an equivalent description of the game implementation in terms of three phases.
In Phase~1, the source (Alice) distributes the correlation resources, $M_{1\to 2}$ and $M_{1\to 3}$,  between the players (Bob and Charlie). 
In Phase~2, the referee generates the question pair $(x,y)$ according to $p_{XY}$, and sends $x$ and $y$ to Bob and Charlie, respectively. 
In Phase~3, upon receiving their questions, Bob and Charlie produce their answers, $B$ and $C$. Once the referee is informed, he decides whether the game is won. 
We refer to this description as a single shot game. %

The winning probability is thus
\begin{align}
\pi(P_{BC|XY})=\sum_{(x,y,b,c)\in\mathcal{X}\times\mathcal{Y}\times\mathcal{B}\times\mathcal{C}} p_{XY}(x,y)P_{BC|XY}(b,c|x,y) \cdot V(x,y,b,c)\,.
\label{Equation: Winning probability}
\end{align}
The performance depends directly on the correlation $P_{BC|XY}$ that Alice, Bob, and Charlie simulate as a consequence of the three phases above. 
 For example, in the CHSH game, %
the winning condition is
$%
    x \wedge y = b \oplus c
$, where $x,y,b,c\in \{0,1\}$, and the winning probability is given by %
$\pi(P_{BC|XY})=\frac{1}{2}\left(1+\frac{S}{4}\right)$.
 Classical strategies may generate correlations $P_{BC|XY}$ such that  $S\leq 2$ (see \eqref{Equation: S quantity}), hence, the game can be won with probability  $\pi(P_{BC|XY})\leq 0.75$. Whereas, entanglement allows for $S=2\sqrt{2}$, for which $\pi(P_{BC|XY})= 0.8535$. 

\subsection{Game realization}
\begin{figure}[t]
    \centering
    \begin{minipage}[b]{0.3\linewidth} %
        \centering
        \includegraphics[scale=0.65,trim={5.3cm 0cm 5.5cm 0}]{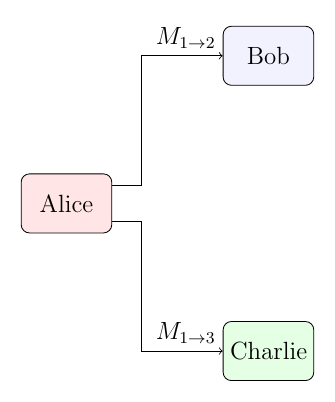} %
        \vspace{0.25cm}
        \begin{center}
        \footnotesize{(a) Phase~1.} %
        \end{center}
    \end{minipage}
    \hspace{0.5cm}%
    \begin{minipage}[b]{0.3\linewidth} %
        \centering
        \includegraphics[scale=0.65,trim={5.3cm 0 5.5cm 0}]{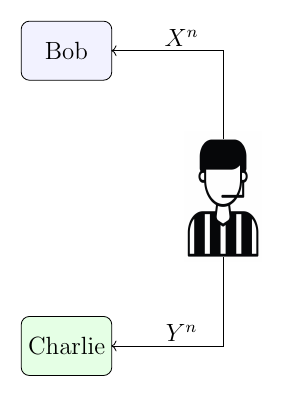} %
        \vspace{0.25cm}
        \begin{center}
        \footnotesize{(b) Phase~2.}%
        \end{center}
    \end{minipage}
    \hspace{0.5cm}%
    \begin{minipage}[b]{0.3\linewidth} %
        \centering
        \includegraphics[scale=0.65,trim={5.3cm 0 5.5cm 0}]{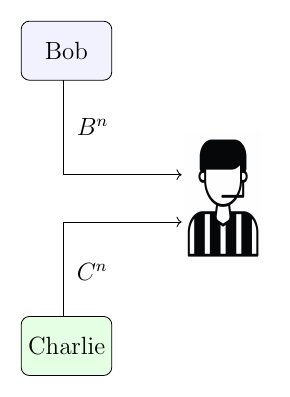} %
        \vspace{0.25cm}
        \begin{center}
        \footnotesize{(c) Phase~3.}%
        \end{center}
    \end{minipage}
    \caption{Implementation of a refereed game in three phases:  %
(a) 
In Phase~1, the source (Alice) distributes the correlation resources  between the players (Bob and Charlie). 
(b) 
In Phase~2, the referee generates the questions, and sends  the respective question to Bob and Charlie.
(c)
In Phase~3, upon receiving their questions, Bob and Charlie produce their answers and inform the referee. Once the referee is informed, he decides whether the game is won.  %
}
    \label{Figure: Nonlocal games}
\end{figure}
In practice, quantum mechanics predict empirical averages.
Therefore, we consider a sequential version, %
see Figure~\ref{Figure: Nonlocal games}. 
In Phase~1, the source (Alice) distributes the correlation resources, $M_{1\to 2}$ and $M_{1\to 3}$,  between the players (Bob and Charlie).  
In   Phase~2, the referee generates a sequence of $n$ independent question pairs $(x_i,y_i)$ according to $p_{XY}$,  and sends $x^n$ and $y^n$ to Bob and Charlie, respectively. 
In Phase~3, the players produce their responses. Bob and Charlie choose their measurements depending on  $x^n$ and $y^n$, respectively.
Then, they perform  their respective measurements on $M_{1\to 2}$ and $M_{1\to 3}$. They   send the measurement outcomes $b^n$ and $c^n$, respectively,  to the referee.
Let $q_{X^n Y^n B^n C^n}$ denote the resulting distribution.
The average winning statistics are characterized by 
\begin{align}
\langle V\rangle %
&=\mathbb{E}\left[\frac{1}{n}\sum_{i=1}^n  V(X_i,Y_i,B_i,C_i) \right]
\nonumber\\
&=\frac{1}{n}\sum_{i=1}^n \sum_{(x_i,y_i,b_i,c_i)\in\mathcal{X}\times\mathcal{Y}\times\mathcal{B}\times\mathcal{C}} \Pr\left( X_i=x_i,Y_i=y_i,B_i=b_i, C_i=c_i\right) \cdot V(x_i,y_i,b_i,c_i)
 \nonumber\\ 
&=\frac{1}{n}\sum_{i=1}^n \left[ \sum_{(x,y,b,c)\in\mathcal{X}\times\mathcal{Y}\times\mathcal{B}\times\mathcal{C}} p_{X Y}(x,y)q_{B_i C_i|X_i Y_i}(b,c|x,y) \cdot V(x,y,b,c) \right]
 \nonumber\\ 
&= \sum_{(x,y,b,c)\in\mathcal{X}\times\mathcal{Y}\times\mathcal{B}\times\mathcal{C}} p_{X Y}(x,y)\bar{q}_{B C|X Y}(b,c|x,y) \cdot V(x,y,b,c)
\end{align}
{where}
\begin{align}
\bar{q}_{B C|X Y}(b,c|x,y)&=\frac{1}{n}\sum_{i=1}^n q_{B_i C_i|X_i Y_i}(b,c|x,y)
\,.
\end{align}
Notice that if we simulate an average correlation of
$\bar{P}_{BC|XY}\approx P_{BC|XY}$, then
\begin{align}
\langle V \rangle\approx \pi(P_{BC|XY}) \,.
\end{align}

\subsection{Coordination as part of a game strategy}
\label{Subsection: Coordination in games}

We now present the connection between  quantum coordination and refereed games explicitly. We  insert a broadcast empirical coordination scheme into the game strategy. Consider the special case where $B$ and $C$ are classical, while
$A$ is null (say, 
$\mathrm{dim}(\mathcal{H}_A)=1$).

Consider the sequential game setup described in Figure~\ref{Figure: Nonlocal games}. 
In Phase 1, the source (Alice) prepares the quantum resources $M_{1\to 2}$ and $M_{1\to 3}$ using the coordination encoding map 
$\mathcal{E}$. She then distributes the resources 
between the respective players (Bob and Charlie), using noiseless quantum links at rates $Q_{1\to 2}$ and $Q_{1\to 3}$. 
In Phase 2, the referee chooses question sequences $X^n$ and $Y^n$ that have no correlation with the quantum resources, as in %
the broadcast network model. %
In Phase 3, Bob and Charlie use the encoding measurements $\mathcal{F}_{X^n M_{1\rightarrow 2}\to B^n}$ and
$\mathcal{D}_{Y^n M_{1\rightarrow 3}\to C^n}$. 
They obtain $B^n$ and $C^n$ as measurement outcomes and inform the referee. 

This coordination strategy generates a classical-correlation state,
${\rho}_{X^n Y^n B^n C^n}$, with an empirical average of ${\overline{\rho}}_{X Y B C}\approx \omega_{XYBC}$, where
\begin{align}
\omega_{XYBC}=\sum_{(x,y,b,c)\in\mathcal{X}\times\mathcal{Y}\times\mathcal{B}\times\mathcal{C}} p_{XY}(x,y)P_{BC|XY}(b,c|x,y)
\ketbra{x,y,b,c} \,,
\end{align}
which leads to a  winning probability $\pi(P_{BC|XY})$ (see \eqref{Equation: Winning probability}).

Let $\mathscr{S}(\gamma)$ denote the set of correlations $P_{BC|XY}$ that induce an average winning probability of at least $\gamma$ in the game.
According to our results,   an average  probability of $\gamma$ is achievalbe if and only if
Alice can transmit qubits to Bob and Charlie at rates $Q_{1\rightarrow 2}$ and $Q_{1\rightarrow 3}$, respectively, such  that the constraints in Theorem~\ref{Theorem: Broadcast - Quantum links} are met some correlation $P_{BC|XY}\in \mathscr{S}(\gamma)$.
Being able to generate entanglement between Bob and Charlie, can provide an advantage by inducing quantum correlations stronger than their classical counterparts, hence allowing for higher winning probabilities (see Subsection~\ref{Subsection: Refereed games}).

\subsection{Example: The CHSH game}
\label{Subsection: CHSH}
A well known game demonstrating the advantage of nonlocal correlations is the CHSH game. Suppose that the players first simulate the following state using a broadcast coordination code: %
\begin{align}
    \ket{\omega^{(x,y)}%
    }&=\sqrt{\alpha_{x,y}}\ket{00}+\sqrt{1-\alpha_{x,y}}\ket{11}\,,
\end{align}
where $\alpha_{x,y}$ are given parameters in $[0,1]$, for 
$(x,y)\in \mathcal{X}\times\mathcal{Y}$. Applying the same measurement strategy as in the CHSH experiment in Subsection~\ref{Subsection: Refereed games}, we obtain a correlation $P$  such that 
the winning probability is given by
\begin{align}
    \pi^{\text{CHSH}}(P)=\frac{1}{16}\sum_{x,y\in \{0,1\}}\left[\frac{1+2\sqrt{2}}{\sqrt{2}}+\sqrt{2\alpha_{x,y}(1-\alpha_{x,y})}\right]\,.
\end{align}
\begin{figure}[t]
    \centering
    \includegraphics[scale=0.22,trim={5.3cm 0 5.5cm 0}]{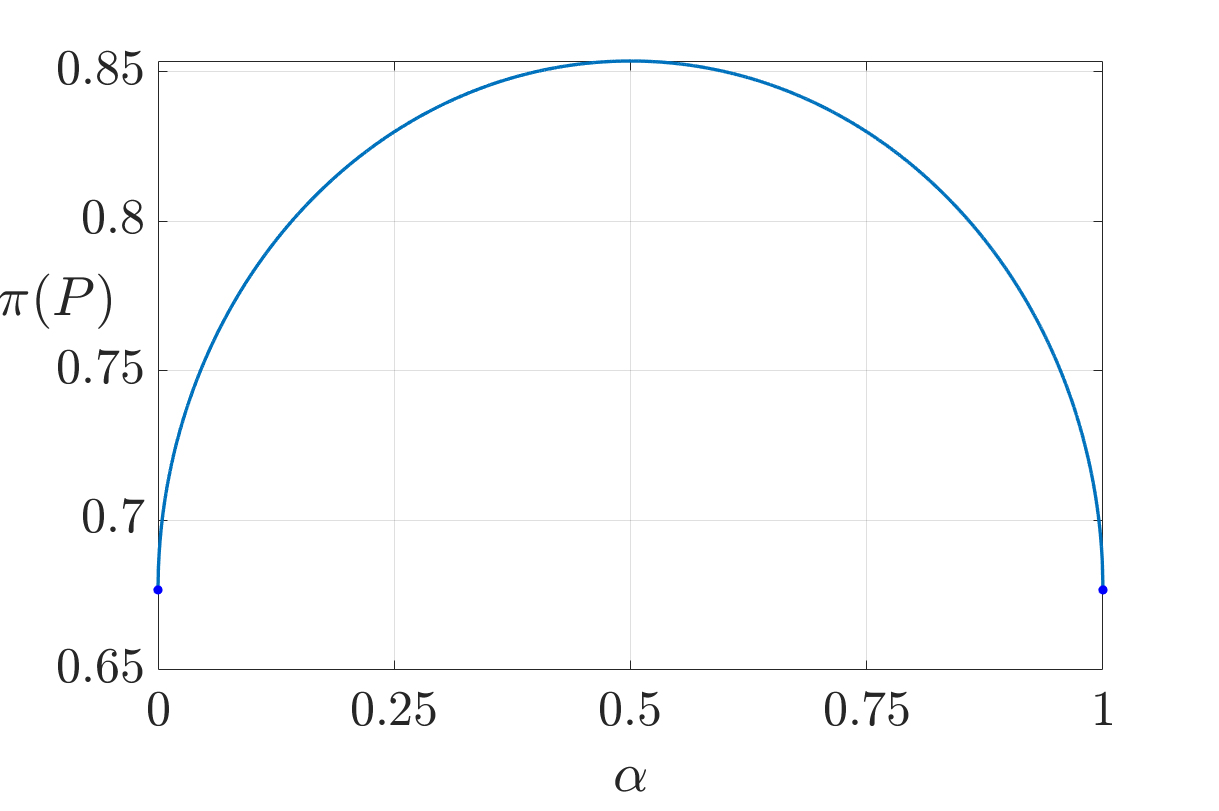}
    \centering
    \caption{ Winning probability as a function of $\alpha$.}  %
    \label{Figure: Winning probability}
\end{figure}
For
$\alpha\equiv \frac{1}{2}$, i.e., with maximal Bell violation \cite{cirel1980quantum}, we get an average winning probability of $\pi^{\text{CHSH}}(P)%
= 0.8535$.
For 
$\alpha\equiv 0$, when there is no correlation, %
the CHSH measurement strategy is even worse that the best classical strategy. %

By Theorem~\ref{Theorem: Broadcast - Quantum links}, Phase 1 requires the communication rates $Q_{1\rightarrow 2}\geq %
\frac{1}{2}h\left( 
 \frac{1}{2}\left( \alpha_{0,0}+\alpha_{0,1} \right)\right)+\frac{1}{2}h\left( 
 \frac{1}{2}\left( \alpha_{1,0}+\alpha_{1,1} \right)\right)$ and $Q_{1\rightarrow 3}\geq %
 \frac{1}{2}h\left( 
 \frac{1}{2}\left( \alpha_{0,0}+\alpha_{1,0} \right)\right)+\frac{1}{2}h\left( 
 \frac{1}{2}\left( \alpha_{0,1}+\alpha_{1,1} \right)\right)$, where $h(\cdot)$ is the binary entropy function. In particular, for a constant parameter, $\alpha_{x,y}=\alpha$ for all $ x,y$, we have $Q_{1\to j}\geq h(\alpha)$. 
There is a threshold value $\alpha^*$ for which the CHSH measurement strategy has the same performance as the best classical strategy. 
Specifically, we obtain a Bell violation provided that $\alpha_{x,y}>0.04491$ for all $(x,y)\in\mathcal{X}\times\mathcal{Y}$.
The winning probability for a constant parameter $\alpha$, is shown in Figure~\ref{Figure: Winning probability}. Since the gradient is unbounded near %
$\alpha=0$, %
even a small amount of entanglement can have a significant effect on the winning probability. As we approach %
$\alpha=\frac{1}{2}$, the gradient diminishes. %
 The Bell violation threshold requires
$Q_{1\to j}\geq h(\alpha^*)= 0.2643$.

To summarize, we have discussed %
the notion of Bell experiments and their direct connection to 
the realization of nonlocal correlations. 
We then discussed refereed games in the standard single-shot form and the sequential form. We have shown that coordination in the broadcast network can be viewed as the overall game strategy, i.e., the preparation of the pre-shared resources (Phase 1) and the measurement (Phase 3). In this sense, empirical coordination is the enabler of quantum strategies that achieve higher  %
 winning statistics compared to classical ones.

\section{Achievability Analysis  (Classical Links)}%
\label{Section: Achievability analysis}

\begin{figure}[tb]
\center
\includegraphics[scale=0.75,trim={5.3cm 0 5.5cm 0}]
{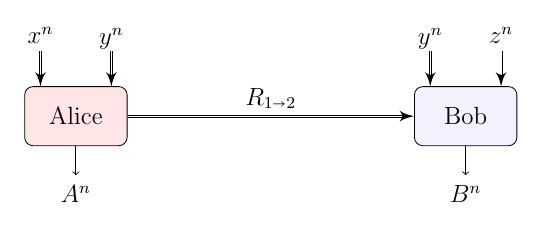} %
\caption{Generic two-node network with a single classical link.}
\label{Figure: generic_lemma}
\end{figure} 
To show the direct part of our coordination capacity theorems, we will use the generic lemma below. 
Consider the generic two-node network in Figure~\ref{Figure: generic_lemma}, where 
Alice receives  $x^n$ and $y^n$ as input to her encoder and
 encodes
a quantum  system $A^n$.
Whereas, Bob receives  $y^n$ and $z^n$ as input and
 encodes a quantum system $B^n$.
 In this case, Alice has encoding maps of the form $p_{M_{1 \shortrightarrow 2}|X^n Y^n S }$ and 
 $\mathcal{E}_{X^n  Y^n S \to A^n}$, and Bob encodes by  
 $\mathcal{F}_{M_{1 \shortrightarrow 2} Y^n Z^n  S\to B^n}$.
 The resulting average state is 
     \begin{align}
        \overline{\rho}_{AB}(x^n,y^n,z^n,s)=\frac{1}{n}\sum_{i=1}^{n}
        \sum_{m_{1 \shortrightarrow 2}\in[2^{n{R_{1 \shortrightarrow 2}}}]} p_{M_{1 \shortrightarrow 2}|S}(m_{1 \shortrightarrow 2}|s) \rho_{A_i}^{(m_{1 \shortrightarrow 2},x^n,y^n,s)}\otimes \rho_{B_i}^{(m_{1 \shortrightarrow 2},y^n,z^n,s)} 
        \,,
    \end{align}
    where $\rho_{A^n}^{(x^n,y^n,s)}=\mathcal{E}_{ X^n Y^n S\to A^n}(x^n,y^n,s)$ and
    $\rho_{B^n}^{(m_{1 \shortrightarrow 2},y^n,z^n,s)}=\mathcal{F}_{M_{1 \shortrightarrow 2}  Y^n Z^nS\to B^n}(m_{1 \shortrightarrow 2},y^n,z^n,s)$.

\begin{lemma}
\label{Lemma: Generic lemma}
 
Consider a state ensemble,
$\{p_{XYZ}\,p_{U|XY} \,,\; \sigma_A^{x,y}\otimes \sigma_B^{y,z,u}\}$.
Let
\begin{align}
\eta_{B}^{x,y,z}= \sum_{u\in\mathcal{U}} p_{U|XY}(u|x,y)  \sigma_B^{y,z,u} \,.
\end{align}
For every $\delta>0$, if 
\begin{align}
   R_{1 \shortrightarrow 2}>I(X;U|YZ)_\sigma
   \,,
\end{align}
then there exists a sequence of randomized $(2^{nR_{1 \shortrightarrow 2}},n)$ empirical coordination codes such that
\begin{align}
  \lim_{n\to \infty}  
  \Pr\left(\norm{
  \overline{\rho}_{AB}(x^n,y^n,z^n,S)-
  \frac{1}{n}
\sum_{i=1}^n 
  \sigma_A^{x_i,y_i}\otimes \eta_B^{x_i,y_i,z_i}
  }_1>\gamma(\delta) 
  \right)=0\,,
\end{align}
uniformly for all $(x^n,y^n,z^n)\in T_{\delta}^{(n)}(p_{XYZ}) $, 
where the probability is computed with respect to the CR element $S$, and $\gamma(\delta)$ tends to zero as $\delta \to 0$.
\end{lemma}

\subsection{Generic scheme: Achievability proof for Lemma~\ref{Lemma: Generic lemma}}
\label{Subsection: Achievability - Geniric lemma}

The proof for Lemma~\ref{Lemma: Generic lemma} is provided below.
Consider the extended c-q state, 
\begin{align}
\sigma_{X Y Z U A B}=
\sum_{(x,y,z,u)\in\mathcal{X}\times \mathcal{Y}\times \mathcal{Z}\times \mathcal{U}}
p_{XYZ}(x,y,z)
p_{U|XY}(u|x,y)\ketbra{x,y,z,u}\otimes
\sigma_A^{x,y}\otimes
\sigma_B^{y,z,u}
\end{align}
where $X$, $Y$, $Z$, and $U$ are classical registers. 
We note that $Z\Cbar (X,Y)\Cbar U$ forms a Markov chain. 

By Theorem~\ref{Theorem: CR theorem}, we may assume that Alice and Bob share unlimited CR. 
Therefore, they can generate the codebook jointly using their random element. 

\paragraph*{Classical codebook construction}
Select $2^{nR_0}$ sequences
$u^n(\ell)$, $\ell\in \big[2^{nR_0}\big]$, independently at random, each i.i.d. according to $p_U$, where
\begin{align}
p_U(u)=\sum_{x,y,z} 
p_{XYZ}(x,y,z) p_{U|XY}(u|x,y)
\,.
\end{align}
Assign each sequence with a bin index $b\left( u^n(\ell) \right)$, where $b:\mathcal{U}^n\to
\big[2^{nR_{1 \shortrightarrow 2}}\big]$, independently at random.
 We thus identify the CR element $S$ as the random codebook $\{u^n(\cdot),b(\cdot)\}$.

\paragraph*{Encoding} 
First, consider the classical encoding function $\mathcal{M}_{1\shortrightarrow2}:\mathcal{X}^n\times
\mathcal{Y}^n\to \big[2^{nR_{1 \shortrightarrow 2}}\big]$. 
Given a pair $(x^n,y^n)\in \mathcal{X}^n\times
\mathcal{Y}^n $, find an index 
$\ell\in \big[2^{nR_0}\big]$ such that 
$(x^n,y^n,u^n(\ell))\in 
T_{2\delta}^{(n)}(p_{XYU})$.
If there is none, set $\ell=1$.
If there is more than one, choose the smallest. 
Send the corresponding bin index, i.e.,
$m_{1 \shortrightarrow 2}(x^n,y^n)=b\left( u^n(\ell) \right)$.

 Then, prepare 
\begin{align}
\rho_{A^n}^{x^n,y^n}
\equiv 
\bigotimes_{i=1}^n \sigma_{A}^{x_i,y_i}
\end{align}

\paragraph*{Decoding}
Given $(y^n,z^n)$ and $m_{1 \shortrightarrow 2}$, find an index $\hat{\ell}\in \big[2^{nR_0}\big]$
such that 
\begin{align}
(y^n,z^n,u^n(\hat{\ell}))\in T_{8\delta}^{(n)}(p_{YZU}) 
\text{ and }
b\left( u^n(\hat{\ell}) \right)=m_{1 \shortrightarrow 2}
\,.
\end{align}
If there is none, set $\hat{\ell}=1$.
If there is more than one, choose the smallest. Prepare the state
\begin{align}
\rho^{y^n,z^n,u^n(\hat{\ell})}_{B^n}
\equiv 
\bigotimes_{i=1}^n \sigma^{y_i,z_i,u_i(\hat{\ell})}_{B}
\end{align}

This results in an average state, 
\begin{align}
\overline{\rho}_{A B}(u^n, x^n,y^n,z^n ) &=
\frac{1}{n} \sum_{i=1}^n 
\rho_{A_i}^{x^n,y^n} \otimes \rho^{y^n,z^n,u^n}_{B_i}
\nonumber\\
&=
\frac{1}{n} \sum_{i=1}^n 
\sigma_{A}^{x_i,y_i}\otimes %
\sigma^{y_i,z_i,u_i}_{B^n}\,,
\end{align}
with $u^n\equiv u^n(\hat{\ell})$.

\paragraph*{Error analysis}
Given $U^n(\hat{\ell})=u^n$, we have
\begin{align}
\overline{\rho}_{A B}(u^n, x^n,y^n,z^n )
&=
\frac{1}{n} \sum_{(a,b,c,d)\in\mathcal{X}\times\mathcal{Y}\times\mathcal{Z}\times\mathcal{U}}
\;
\sum_{i: (x_i,y_i,z_i,u_i)=(a,b,c,d)} \sigma_A^{x_i,y_i}\otimes\sigma^{y_i,z_i,u_i}_{B}
\nonumber\\
&=
\frac{1}{n} \sum_{(a,b,c,d)\in\mathcal{X}\times\mathcal{Y}\times\mathcal{Z}\times\mathcal{U}}
\;
\sum_{i: (x_i,y_i,z_i,u_i)=(a,b,c,d)} \sigma_A^{a,b}\otimes\sigma^{b,c,d}_{B}
\nonumber\\
&=
\sum_{(a,b,c,d)\in\mathcal{X}\times\mathcal{Y}\times\mathcal{Z}\times\mathcal{U}}
\hat{P}_{x^n,y^n,z^n,u^n}(a,b,c,d)  \sigma_A^{a,b}\otimes \sigma^{b,c,d}_{B}
\nonumber\\
&=
\sum_{(a,b,c)\in\mathcal{X}\times\mathcal{Y}\times\mathcal{Z}}
\hat{P}_{x^n,y^n,z^n}(a,b,c) 
\sum_{d\in\mathcal{U}}
\hat{P}_{u^n|x^n,y^n,z^n}(d|a,b,c)  \sigma_A^{a,b}\otimes \sigma^{b,c,d}_{B} \,.
\end{align}

For every $u^n$ such that 
$(x^n,y^n,z^n,u^n)\in T^{(n)}_{\gamma(\delta)}(p_{XYZU})$,
\begin{align}
\norm{\overline{\rho}_{A B}(u^n, x^n,y^n,z^n )-  \tau_{AB} }_1 &\leq \gamma(\delta)
\end{align}
where
\begin{align}
\tau_{AB}&=
\sum_{(a,b,c)\in\mathcal{X}\times\mathcal{Y}\times\mathcal{Z}}
\hat{P}_{x^n,y^n,z^n}(a,b,c) 
\sum_{d\in\mathcal{U}}
p_{U|XYZ}(d|a,b,c)  \sigma_A^{a,b}\otimes \sigma^{b,c,d}_{B}
\nonumber\\
&=
\sum_{(a,b,c)\in\mathcal{X}\times\mathcal{Y}\times\mathcal{Z}}
\hat{P}_{x^n,y^n,z^n}(a,b,c) 
  \sigma_A^{a,b}\otimes \sum_{d\in\mathcal{U}}
p_{U|XY}(d|a,b)\sigma^{b,c,d}_{B}
\nonumber\\
&=
\sum_{(a,b,c)\in\mathcal{X}\times\mathcal{Y}\times\mathcal{Z}}
\hat{P}_{x^n,y^n,z^n}(a,b,c) 
  \sigma_A^{a,b}\otimes \eta^{a,b,c}_{B}
\nonumber\\
&=
\frac{1}{n}
\sum_{(a,b,c)\in\mathcal{X}\times\mathcal{Y}\times\mathcal{Z}}
N(a,b,c|x^n,y^n,z^n) 
  \sigma_A^{a,b}\otimes \eta^{a,b,c}_{B}
  \nonumber\\
&=
\frac{1}{n}
\sum_{(a,b,c)\in\mathcal{X}\times\mathcal{Y}\times\mathcal{Z}}
\sum_{i: (x_i,y_i,z_i)=(a,b,c)} 
  \sigma_A^{a,b}\otimes \eta^{a,b,c}_{B}
  \nonumber\\
&=
\frac{1}{n}
\sum_{(a,b,c)\in\mathcal{X}\times\mathcal{Y}\times\mathcal{Z}}
\sum_{i: (x_i,y_i,z_i)=(a,b,c)} 
  \sigma_A^{x_i,y_i}\otimes \eta_B^{x_i,y_i,z_i}
    \nonumber\\
&=
\frac{1}{n}
\sum_{i=1}^n 
  \sigma_A^{x_i,y_i}\otimes \eta_B^{x_i,y_i,z_i} \,.
\end{align}

Consider the event
\begin{align}
\mathcal{A}_1\equiv 
\left\{ (x^n,y^n,z^n,U^n(\hat{j}))\in T^{(n)}_{\gamma(\delta)}(p_{XYZU}) \right\}
\label{Equation:jointly_typical}
\end{align}
Based on the classical result due to Cuff et al.
\cite[Lemm. 14]{cuff2010coordination},
\begin{align}
\Pr(\mathcal{A}_1)\geq
1-\alpha_n
\end{align}
for all $(x^n,y^n,z^n)\in T^{(n)}_\delta(p_{XYZ})$,
where $\gamma\equiv \gamma(\delta)$ tends to zero as 
$\delta\to 0$, and $\alpha_n$ tends to zero as $n\to\infty$, provided that
\begin{align}
R &> I(X;U|YZ)+\gamma(\delta) \,.
\end{align}
Therefore,
\begin{align}
\Pr\left(
\norm{\overline{\rho}_{A B}(U^n(\hat{\ell}), x^n,y^n,z^n )-  \tau_{AB} }_1 >\gamma(\delta) \right)
\leq \alpha_n
\,.
 \end{align}
This completes the proof of Lemma~\ref{Lemma: Generic lemma}.

We are now in a position to give the achievability proofs for the two-node and cascade networks.
\subsection{Two-node network: Achievability proof for Theorem~\ref{Theorem: Two nodes}}
\label{Achievability: Two nodes}
The proof essentially follows from Lemma~\ref{Lemma: Generic lemma}, with the following addition.
If Alice receives a random sequence $X^n$ that is not $\delta$-typical, then she sends an arbitrary transmission. Otherwise, she encodes using the encoder in Lemma~\ref{Lemma: Generic lemma}.
Since $\Pr\left(X^n\in T_\delta^{(n)}(p_X)\right)$ tends to $1$ as $n\to\infty$,
 achievability  for the two-node network follows.

\subsection{Cascade network: Achievability proof for Theorem~\ref{Theorem: Cascade}}
\label{Achievability: Cascade}
We use rate splitting, where Alice's message consists of two components
$m_{1 \shortrightarrow 2}'$ and $m_{1 \shortrightarrow 2}''$, at rates 
$R_{1 \shortrightarrow 2}'$ and $R_{1 \shortrightarrow 2}''$, respectively, where
$R_{1 \shortrightarrow 2}=R_{1 \shortrightarrow 2}'+R_{1 \shortrightarrow 2}''$.

\paragraph*{Classical codebook construction}
Select $2\cdot 2^{nR_0}$ sequences
$y^n(\ell')$, $z^n(\ell'')$, $\ell',\ell''\in \big[2^{nR_0}\big]$, independently at random, each i.i.d. according to $p_Y$ and $p_Z$, where
\begin{align}
p_{YZ}(y,z)=\sum_{x} 
p_{X}(x) p_{YZ|X}(y,z|x)
\,.
\end{align}
Assign each sequence with a bin index $b\left( y^n(\ell') \right)$ and $c\left( z^n(\ell'') \right)$, where $b:\mathcal{Y}^n\to
[2^{nR_{1 \shortrightarrow 2}'}]$ and $c:\mathcal{Z}^n\to
[2^{nR_{1 \shortrightarrow 2}''}]$, independently at random.

\paragraph*{Alice's encoder} 
As before, if Alice receives  $x^n\notin T_\delta^{(n)}(p_X)$,  she sends an arbitrary transmission.
Otherwise, consider the classical encoding function $\mathcal{M}_{1\shortrightarrow2}:\mathcal{X}^n\to [2^{nR_{1 \shortrightarrow 2}'}]\times [2^{nR_{1 \shortrightarrow 2}''}]$ below. 
Given  $x^n\in T_\delta^{(n)}(p_X)$, find an index pair 
$(\ell',\ell'')\in \big[2^{nR_0}\big]\times \big[2^{nR_0}\big]$ such that 
$(x^n,y^n(\ell'),z^n(\ell''))\in 
T_{2\delta}^{(n)}(p_{XYZ})$.
If there is none, set $(\ell',\ell'')=(1,1)$.
If there is more than one, choose the first. 
Send the corresponding bin indices, i.e.,
$m_{1 \shortrightarrow 2}'(x^n)=b\left( y^n(\ell') \right)$ and $m_{1 \shortrightarrow 2}''(x^n)=c\left( z^n(\ell'') \right)$.

 Then, prepare 
\begin{align}
\rho_{A^n}^{x^n}
\equiv 
\bigotimes_{i=1}^n \sigma_{A}^{x_i}
\,.
\end{align}

\paragraph*{Bob's encoder}
Bob receives $m_{1 \shortrightarrow 2}=(m_{1 \shortrightarrow 2}', m_{1 \shortrightarrow 2}'')$, and  encodes in three stages:
\begin{enumerate}[(i)]
\item
Given  $m_{1 \shortrightarrow 2}''$, find an index $\hat{\ell}''\in \big[2^{nR_0}\big]$
such that 
\begin{align}
z^n(\hat{\ell}'')\in T_{8\delta}^{(n)}(p_{Z}) 
\text{ and }
c\left( z^n(\hat{\ell}'') \right)=m_{1 \shortrightarrow 2}''
\,.
\end{align}
If there is none, set $\hat{\ell}''=1$.
If there is more than one, choose the smallest. 
Send $m_{2 \shortrightarrow 3}=m_{1 \shortrightarrow 2}''$ to Charlie.

\item
Now given  $m_{1 \shortrightarrow 2}'$ and $\hat{\ell}''$, find an index $\hat{\ell}'\in \big[2^{nR_0}\big]$
such that 
\begin{align}
(y^n(\hat{\ell}'),z^n(\hat{\ell}''))\in T_{8\delta}^{(n)}(p_{YZ}) 
\text{ and }
b\left( y^n(\hat{\ell}') \right)=m_{1 \shortrightarrow 2}'
\,.
\end{align}
If there is none, set $\hat{\ell}'=1$.
If there is more than one, choose the smallest. 

\item
Prepare the state
\begin{align}
\rho^{y^n(\hat{\ell}')}_{B^n \bar{Z}^n}
\equiv 
\bigotimes_{i=1}^n \sigma^{y_i(\hat{\ell}')}_{B}\otimes 
\ketbra{z_i(\hat{\ell}'')}_{\bar{Z}}
\end{align}
where $\bar{Z}^n$ is an auxiliary system for Bob.
\end{enumerate}

\paragraph*{Charlie's encoder}
Given  $m_{2 \shortrightarrow 3}=m_{1 \shortrightarrow 2}''$, find an index $\tilde{\ell}''\in \big[2^{nR_0}\big]$
such that 
\begin{align}
z^n(\tilde{\ell}'')\in T_{8\delta}^{(n)}(p_{Z}) 
\text{ and }
c\left( z^n(\tilde{\ell}'') \right)=m_{1 \shortrightarrow 2}''
\,.
\end{align}
If there is none, set $\tilde{\ell}''=1$.
If there is more than one, choose the smallest.

Prepare the state
\begin{align}
\rho^{z^n\left(\tilde{\ell}''\right)}_{C^n}
\equiv 
\bigotimes_{i=1}^n \sigma^{z_i\left(\tilde{\ell}''\right)}_{C}
\end{align}
This results in an average state, 
\begin{align}
\overline{\rho}_{A B\bar{Z} C}( x^n,y^n,z^n ) &=
\frac{1}{n} \sum_{i=1}^n 
\rho_{A_i}^{x^n} \otimes \rho^{y^n}_{B_i}
\otimes 
\ketbra{\bar{z}^n}
\otimes
\rho^{z^n}_{C_i}
\nonumber\\
&=
\frac{1}{n} \sum_{i=1}^n 
\sigma_{A}^{x_i}\otimes %
\sigma^{y_i}_{B}\otimes
\ketbra{\bar{z}_i}\otimes
\sigma^{z_i}_{C}\,,
\end{align}
with $y^n\equiv y^n(\hat{\ell}')$,
$\bar{z}^n\equiv z^n(\hat{\ell}'')$,
and 
$z^n\equiv z^n(\tilde{\ell}'')$.
Based on the analysis in the proof of Lemma~\ref{Lemma: Generic lemma} (see Section~\ref{Subsection: Achievability - Geniric lemma}), Alice, Bob, and Charlie achieve empirical coordination of  
$\sigma_{ABZC}$, provided that 
\begin{align}
R_{2 \shortrightarrow 3}=R_{1 \shortrightarrow 2}''&>I(X;Z)
\,,
\\
R_{1 \shortrightarrow 2}'&>I(X;Y|Z)
\end{align}
which requires
$R_{1 \shortrightarrow 2}=R_{1 \shortrightarrow 2}'+R_{1 \shortrightarrow 2}''>I(X;YZ)$.
\qed

\section{Converse Part Analysis (Classical Links)}%
\label{Section: Converse analysis}
We now show the converse part for our coordination capacity theorems. 

\subsection{Two-node network: Converse proof for Theorem~\ref{Theorem: Two nodes}}
\label{Converse: Two nodes}
Consider the two-node network in Figure~\ref{Figure: Two nodes}.
Let $R_{1 \shortrightarrow 2}$ be an achievable rate  for empirical coordination with a desired state $\omega_{AB}$. Then, there exists a sequence of $\left(2^{n{R_{1 \shortrightarrow 2}}},n\right)$ empirical coordination codes that achieves an error, 
\begin{align}
\label{Error: Two nodes}
\norm{\overline{\rho}_{XAB}-\omega_{XAB}}_1 \leq \varepsilon_n  \,,
\end{align} 
where $\varepsilon_n$ tends to zero as $n\to\infty$. 
Now, suppose that Bob  performs a projective measurement in a particular basis, say,  $\{\ket{y}\}$.
This yields a sequence $Y^n$ as the measurement outcome, with 
some distribution $p_{ Y^n|X^n}(y^n|x^n)$.

Then, consider the classical variables $X_J$ and $Y_J$, where $J$ is a uniformly distributed random variable, over the index set $[n]$, drawn independently of $X^n$, $Y^n$.
Their joint distribution is
\begin{align}
\bar{p}_{X_J Y_J}(x,y)&=
\frac{1}{n}\sum_{i=1}^n p_{X_i Y_i}(x,y)
\nonumber\\&=
\left(\bra{x}\otimes \bra{y}\right)
\overline{\rho}_{XB}\left(\ket{x}\otimes \ket{y}\right)\,,
\end{align}
where $p_{X_i Y_i}$ is the marginal distribution of $p_X^n\times p_{Y^n|X^n}$.
Based on \eqref{Error: Two nodes}, we have the following total variation bound:
\begin{align}
\label{Error: Two nodes distribution}
\norm{\bar{p}_{X_J Y_J}-\pi_{XY}}_1 \leq \varepsilon_n  \,,
\end{align}
where $\pi_{X Y}$ is defined as
\begin{align}
\pi_{X Y}(x,y)&=
\left(\bra{x}\otimes \bra{y}\right)
\omega_{AB}\left(\ket{x}\otimes \ket{y}\right)\,,
\end{align}
for $(x,y)\in\mathcal{X}\times\mathcal{Y}$.

Next, consider that
\begin{align}
nR_{1 \shortrightarrow 2}
&\geq 
H(M_{1 \shortrightarrow 2})
\nonumber\\&\geq 
I(X^n;M_{1 \shortrightarrow 2})
\nonumber\\&\geq 
I(X^n;Y^n)
\nonumber\\&= 
\sum_{i=1}^n I(X_i;Y^n|X^{i-1})
\nonumber\\&= 
\sum_{i=1}^n I(X_i;X^{i-1} Y^n)
\nonumber\\&\geq 
\sum_{i=1}^n I(X_i; Y_i)
\nonumber\\&= 
n I(X_J; Y_J|J)_{\bar{p}}
\label{Equation: Two Node Converse DPI}
\end{align}
where the third inequality holds by the data processing inequality and the following equalities by the chain rule.
Since $X^n$ is i.i.d., it follows that $X_J$ and $J$ are statistically independent, hence, 
\begin{align}
I(X_J; Y_J|J)_{\bar{p}}&=I(X_J; Y_J J)_{\bar{p}}
\nonumber\\
&\geq I(X_J; Y_J )_{\bar{p}} \,.
\end{align}
Based on entropy continuity \cite{shannon1948mathematical},
\begin{align}
I(X_J; Y_J)_{\bar{p}}&\geq  
I(X; Y)_\pi-\alpha_n
\end{align}
where $\alpha_n=-3\varepsilon_n \log(\varepsilon_n\abs{\mathcal{X}}\abs{\mathcal{Y}})$ 
\cite[Lemm. 2.7]{csiszar2011information}, which tends to zero as $n\to\infty$.
This concludes the converse proof for the two-node network.

\subsection{Cascade network: Converse proof for Theorem~\ref{Theorem: Cascade}}
\label{Converse: Cascade}

Consider the cascade network in Figure~\ref{Figure: Cascade network}.
If %
$(R_{1 \shortrightarrow 2},R_{2 \shortrightarrow 3})$ is achievable, then %
there exists a sequence of  $\left(2^{n{R_{1 \shortrightarrow 2}}},2^{n{R_{2 \shortrightarrow 3}}},n\right)$   codes such %
\begin{align}
\label{Error: Cascade}
\norm{\overline{\rho}_{XABC}-\omega_{XABC}}_1 \leq \varepsilon_n  \,,
\end{align} 
where $\varepsilon_n$ tends to zero as $n\to\infty$. 
Suppose that Bob and Charlie perform  projective measurements in a particular basis, say,  $\{\ket{y}\}$ and $\{\ket{z}\}$, respectively.
This yields a sequence $(Y^n,Z^n)$ as the measurement outcomes, with 
some distribution $p_{ Y^n Z^n|X^n}(y^n,z^n|x^n)$.

Then, consider the classical variables $X_J$, $Y_J$, and $Z_J$, where $J$ is uniform over  $[n]$, independent of $X^n$, $Y^n$, and $Z^n$.
Their joint distribution is
\begin{align}
\bar{p}_{X_J Y_J Z_J}(x,y,z)&=
\frac{1}{n}\sum_{i=1}^n p_{X_i Y_i Z_i}(x,y,z)
\nonumber\\&=
(\bra{x}\otimes \bra{y}\otimes \bra{z})
\overline{\rho}_{XBC}(\ket{x}\otimes \ket{y}\otimes \ket{z})\,,
\end{align}
where $p_{X_i Y_i Z_i}$ is the marginal distribution of $p_X^n\times p_{Y^n Z^n|X^n}$.
By \eqref{Error: Cascade}, %
\begin{align}
\label{Error: Cascade distribution}
\norm{\bar{p}_{X_J Y_J Z_J}-\pi_{XYZ}}_1 \leq \varepsilon_n  \,,
\end{align}
where
\begin{align}
\pi_{X Y Z}(x,y,z)&=
(\bra{x}\otimes \bra{y}\otimes \bra{z})
\omega_{ABC}(\ket{x}\otimes \ket{y}\otimes \ket{z})\,,
\end{align}
for $(x,y,z)\in\mathcal{X}\times\mathcal{Y}\times \mathcal{Z}$.

\begin{figure}[tb]
\center
\includegraphics[scale=0.75,trim={5.3cm 0 5.5cm 0}]
{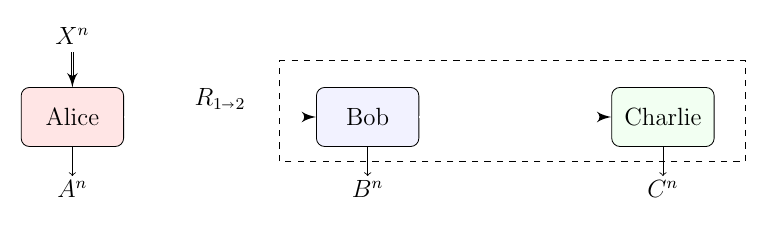} %
\caption{Encoding by Bob and Charlie.}
\label{Figure: Cascade in a box}
\end{figure}

Consider Alice's communication rate, $R_{1 \shortrightarrow 2}$. Now, we may view the overall encoding operation of Bob and Charlie as a ``black box" with $M_{1 \shortrightarrow 2}$ as input and $(B^n,C^n)$ as output, as shown in Figure~\ref{Figure: Cascade in a box}.
Thus, 
\begin{align}
nR_{1 \shortrightarrow 2}
&\geq 
H(M_{1 \shortrightarrow 2})
\nonumber\\&\geq 
I(X^n;M_{1 \shortrightarrow 2})
\nonumber\\&\geq 
I(X^n;Y^n Z^n)
\nonumber\\&= 
\sum_{i=1}^n I(X_i;Y^n Z^n|X^{i-1})
\nonumber\\&= 
\sum_{i=1}^n I(X_i;X^{i-1} Y^n Z^n)
\nonumber\\&\geq 
\sum_{i=1}^n I(X_i; Y_i Z_i)
\nonumber\\&= 
n I(X_J; Y_J Z_J|J)_{\bar{p}}
\end{align}
based on the same arguments as in \eqref{Equation: Two Node Converse DPI}. Since $X_J$ and $J$ are statistically independent, we have
\begin{align}
R_{1 \shortrightarrow 2}&\geq I(X_J; Y_J Z_J|J)_{\bar{p}}
\nonumber\\
&=
I(X_J;JY_J Z_J)_{\bar{p}}
\nonumber\\
&\geq
I(X_J;Y_J Z_J)_{\bar{p}}
\end{align}
Following similar steps, we also have
\begin{align}
R_{2 \shortrightarrow 3}\geq I(X_J;Z_J)_{\bar{p}}
\,.
\end{align}

Based on entropy continuity \cite{shannon1948mathematical},
\begin{align}
I(X_J; Y_J Z_J)_{\bar{p}}&\geq  
I(X; Y Z)_\pi-\alpha_n
\,,\;
\\
I(X_J;  Z_J)_{\bar{p}}&\geq  
I(X;  Z)_\pi-\alpha_n
\end{align}
where  $\alpha_n=-3\varepsilon_n \log(\varepsilon_n\abs{\mathcal{X}}\abs{\mathcal{Y}}\abs{\mathcal{Z}})$ 
\cite[Lemm. 2.7]{csiszar2011information}, which tends to zero as $n\to\infty$.
\qed

\section{Two-Node and Broadcast Analysis  (Quantum Links) }
\label{Section: Analysis - Quantum links}
\subsection{%
Proof of Theorem~\ref{Theorem: Two-node - Quantum links} (two-node network)}
\label{Converse: Two nodes - Empirical - Quantum links}

\begin{figure}[t]
\center
\includegraphics[scale=0.75,trim={3.75cm 0 4cm 0}]
{QL_two_node_network_Empirical.pdf} %
\caption{Two-node network with a quantum link.}
\label{Figure: Two_nodes_Empirical_QL}
\end{figure}
The achievability proof for the two-node network with a quantum link immediately follows from the achievability in the strong coordination framework  \cite{natur2025quantum} (see discussion on strong vs. empirical coordination in Subsection~\ref{Subsection:Strong_vs_Empirical}).
We now focus on the converse part of Theorem~\ref{Theorem: Two-node - Quantum links}.
Alice prepres the state $\ket{\omega_{RA\bar{B}}}^{\otimes n}$ locally, where $\bar{B}$ is a local ancilla. 
Then, she applies an encoding map $\mathcal{E}_{\bar{B}^n  \to M_{1\rightarrow 2}}$ on her part, and sends the output $M_{1\rightarrow 2}$ to Bob.  %
As Bob receives $M_{1\rightarrow 2}$, %
he encodes %
by $\mathcal{F}_{M_{1\rightarrow 2} \to B^{n}}$.
The protocol can be described through the following relations:
\begin{align} 
\tau_{R^n A^{n} M_{1\rightarrow 2}}&=
\nonumber
(\mathrm{id}_{R^nA^n}\otimes \mathcal{E}_{ \bar{B}^n  \to M_{1\rightarrow 2}})
(\omega_{RA\bar{B}}^{\otimes n} ) \,, 
\\
\rho_{R^n A^{n}B^{n}}&=
(\mathrm{id}_{R^n A^{n}} \otimes\mathcal{F}_{M_{1\rightarrow 2}\to B^{n}})
(\tau_{R^n A^{n}M_{1\rightarrow 2} }
)\,.
\end{align}

Let $Q_{1\rightarrow 2}$ be an achievable rate for empirical coordination in the two-node network with respect to $\ket{\omega_{RAB}}$. Then, there exists a sequence of  codes $\left( 2^{nQ_{1\rightarrow 2}},n\right)$ %
such that 
\begin{align}
   \frac{1}{2} \left\Vert \frac{1}{n} \sum_{i=1}^n \rho_{R_i A_{i}B_{i}} - \omega_{RAB} \right\Vert _{1} \leq \varepsilon_n \,,
\label{Two_nodes_Error_Converse_QL}
\end{align}
where $\omega_{RAB}\equiv \ketbra{\omega_{RAB}}$, and
$\varepsilon_n \to 0$ as $n\to \infty$.
Then,
\begin{align}
2nQ_{1\rightarrow 2}	&=2\left[\log\dim(\mathcal{H}_{M_{1\rightarrow 2}})\right]
\nonumber\\
& {\geq}I(M_{1\rightarrow 2};R^{n}A^{n})_{\tau}
\,.
\label{Equation: Converse Two-node Dim(H_M)}
\end{align}
By the data processing inequality and  the chain rule for the quantum mutual information, %
\begin{align}
I(M_{1\rightarrow 2};R^{n}A^{n})_{\tau}
\,&{\geq}I(B^n ;R^{n}A^{n})_{\rho}
\nonumber\\
&=\sum_{i=1}^n I(R_iA_i;B^n|R^{i-1}A^{i-1} )_{\rho}
\,.
\label{Equation: Converse Two-node Chain rule}
\end{align}

Observe that Alice and Bob's encoders do not act on $R^n A^n$.
Therefore, $\rho_{R^n A^n}=\omega_{RA}^{\otimes n}$.
This, in turn, implies
\begin{align}
I(R_iA_i;R^{i-1}A^{i-1} )_{\rho}
&=
I(R_iA_i;R^{i-1}A^{i-1} )_{\omega^{\otimes n}}
\nonumber\\
&=0 \,.
\end{align}
Hence, 
\begin{align}
I(R_iA_i;B^n|R^{i-1}A^{i-1} )_{\rho}%
&= I(R_iA_i;R^{i-1}A^{i-1}B^n )_{\rho}
\nonumber\\
&\geq 
I(R_iA_i;B_i )_{\rho}
\,.
\label{Equation: Converse Two-node IID}
\end{align}

Now, define a random variable $J\sim \text{Uniform}[n]$. We denote denote the extended c-q-q-q state by 
 \begin{align}
 \sigma_{JRAB}=\sum_{j=1}^n \ketbra{j}_J\otimes \rho_{R_jA_jB_j}
 \,.
 \label{Equation:Sigma_J}
 \end{align} 
 Thus, by \eqref{Equation: Converse Two-node Dim(H_M)}, \eqref{Equation: Converse Two-node Chain rule} and \eqref{Equation: Converse Two-node IID},
\begin{align}
Q_{1\to 2}&\geq 
\frac{1}{2n}\sum_{i=1}^n I(R_iA_i;B_i )_{\rho}
\nonumber\\
&
= \frac{1}{2} I(R_JA_J;B_J|J )_{\sigma}\,.
\label{Equation: Converse Two-node Time sharing}
\end{align}

Since Alice begins with preparing a memoryless state, the systems $R_J A_J$ and $J$ are uncorrelated, as
$\sigma_{JR_J A_J}=\frac{\identity_J}{n}\otimes \omega_{RA}$.
Thus, 
\begin{align}
I(R_JA_J;B_J|J)_{\sigma} &=
I(R_JA_J;B_JJ)_{\sigma} 
\nonumber\\
& \geq
 I(R_JA_J;B_J)_{\sigma}\,.
 \label{Equation: Converse Two-node independent J}
\end{align}
Observe that 

\begin{align}
    \frac{1}{2}\left\Vert \sigma_{R_J A_{J}B_{J}} - \omega_{RAB} \right\Vert _{1} \leq \varepsilon_n \,,
\label{Two_nodes_Error_Converse_QL_Sigma}
\end{align}
by \eqref{Two_nodes_Error_Converse_QL} and \eqref{Equation:Sigma_J}.
Therefore, by  entropy continuity \cite{AFW_Winter_2016}
and since $\ket{\omega_{RAB}}$ is pure,%
\begin{align}
 I(R_JA_J;B_J)_{\sigma}
&\geq  I(RA;B )_{\omega} - \beta_n
\nonumber\\
&=  2H(B)_{\omega} - \beta_n\,,
\label{Equation: Converse Two-node AFW}
\end{align}
where $\beta_n=3\varepsilon_n \log \text{dim}(\mathcal{H}_{RA})+2(1+\varepsilon_n)h\left(\frac{\varepsilon_n}{1+\varepsilon_n}\right)$ and $h(\cdot)$ is the binary entropy function introduced in the Section~\ref{Section: Notation}. Note that $\beta_n$ tends to zero as $n\to \infty$.
Hence, the proof follows from \eqref{Equation: Converse Two-node Time sharing}, \eqref{Equation: Converse Two-node independent J}, and \eqref{Equation: Converse Two-node AFW}.

\subsection{Proof of Theorem~\ref{Theorem: Broadcast - Quantum links} %
(broadcast network)}
\label{Section: Analysis - Broadcast - Empirical - Quantum links}

\begin{figure}[t]
\center
\includegraphics[scale=0.75,trim={3.75cm 0 4cm 0}]
{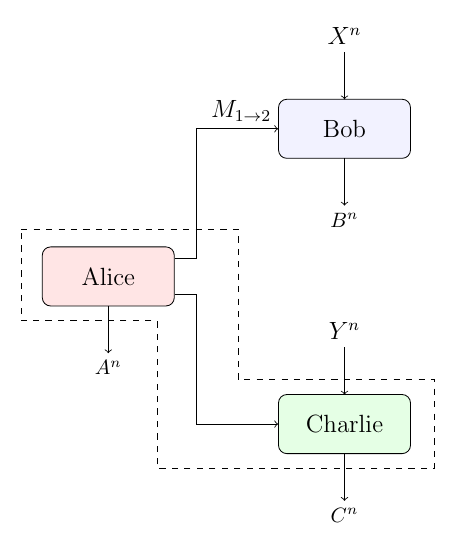} %
\caption{Full cooperation between Alice and Charlie in the broadcast network.}
\label{Figure: Broadcast_Empirical}
\end{figure}

As in the two-node analysis,
 achievability immediately follows from the strong coordination result in \cite[Th. 9]{natur2025quantum}. We now prove the converse part of Theorem~\ref{Theorem: Broadcast - Quantum links}. 
We first begin by proving the lower bound on the communication rate between Alice and Bob, $Q_{1\rightarrow 2}$. 
We observe that for the converse proof, it suffices to show the lower bound while
assuming that Alice and Charlie have full cooperation, therefore, Alice has access to the  $Y^n$ and $C^n$. In this case, one  may think of Alice and Charlie as one entity with a classical input $Y^n$ and  quantum outputs $A^n$, $C^n$, and $M_{1\to 2}$. See Figure~\ref{Figure: Broadcast_Empirical}.

 Define the following state 
\begin{align} 
\omega_{X A B C}^{(y)}&= \sum_{y\in\mathcal{Y} } p_{X|Y}(x|y) \ketbra{x} \otimes \ketbra{\omega^{(x,y)}_{A\bar{B}C}}\,,
\end{align}
Based on our assumption in
\eqref{Equation:Charlie_Non_Signaling}, the reduced state of Alice and Charlie does not depend on $x$, when conditioned on $y$.
That is, $\omega_{A C }^{(x,y)}\equiv \omega_{ A C }^{(y)}$.
Let $\ket{\omega_{ A C R }^{(y)}}$ be a purification of this state. 

Suppose Alice  prepares the state
\begin{align}
\ket{\omega_{\bar{R}^n A^n C^n}^{(y^n)}}\equiv \bigotimes_{i=1}^n \ket{\omega_{\bar{R} A C}^{(y_i)}}
\,.
\end{align}
She applies an encoding map $\mathcal{E}^{(y^n)}_{
\bar{R}^n  \to M_{1\rightarrow 2}}$ on her part, and 
sends the quantum description $M_{1\rightarrow 2}$  to Bob, thereafter he applies his own encoder,
$\mathcal{F}_{  M_{1\rightarrow 2} X^n \to  B^n X^n} $.
The encoding scheme can be described by the %
\begin{align}
\tau_{X^n A^{n}  C^n M_{1\rightarrow 2}}^{(y^n)} &=
\omega_{X^n}^{(y^n)}\otimes \left[( \mathrm{id}_{A^n {C}^n} 
\otimes \mathcal{E}^{(y^n)}_{%
\bar{R}^n  \to M_{1\rightarrow 2}})
(\omega_{ A^n \bar{R}^n C^n}^{(y^n)}) \right] \,, 
\nonumber
\\
\rho_{Y^n  A^{n}C^n B^{n} X^n} &=
\sum_{y^n\in\mathcal{Y}^n} p^{\otimes n}_Y(y^n) \ketbra{y^n}\otimes 
(\mathrm{id}_{A^n C^n}\otimes\mathcal{F}_{  M_{1\rightarrow 2} X^n \to  B^n X^n}  
)
(\tau_{ A^{n}  C^n M_{1\rightarrow 2} X^n}^{(y^n)}) \,. 
\end{align}

Let $Q_{1\rightarrow 2}$ be an achievable rate. Then, there exists a sequence of  codes 
such that 
\begin{align}
    \frac{1}{2}\left\Vert \frac{1}{n} \sum_{i=1}^n \rho_{X_i Y_i A_{i}B_{i} C_i} - \omega_{XYABC} \right\Vert _{1} \leq \varepsilon_n \,,
\label{Broadcast_Error_Converse_QL}
\end{align}
where 
$\varepsilon_n \to 0$ as $n\to \infty$.
Then,
\begin{align}
2nQ_{1\rightarrow 2}	&=2\left[\log\dim(\mathcal{H}_{M_{1\rightarrow 2}})\right]
\nonumber\\
& {\geq}I(M_{1\rightarrow 2};A^{n} {C}^n Y^n|X^n)_{\tau}\,.
\label{Equation: Converse Broadcast Dim(H_M)}
\end{align}
By the data processing inequality and  the chain rule for the quantum mutual information, %
\begin{align}
I(M_{1\rightarrow 2};A^{n}C^n Y^n|X^n)_{\tau}
\,&{\geq}I(B^n ;A^{n} C^n Y^n|X^n)_{\rho}
\nonumber\\
&=\sum_{i=1}^n I(A_i C_i Y_i;B^n|A^{i-1} C^{i-1}Y^{i-1} X^n )_{\rho}
\,.
\label{Equation: Converse Broadcast Chain rule}
\end{align}

Observe that Alice and Bob's encoders do not act on $ Y^nA^nC^n$.
    Therefore, $\rho_{Y^n A^n C^n} = \omega_{YAC}^{\otimes n}$.
This, in turn, implies
\begin{align}
I(A_i C_i Y_i;A^{i-1} C^{i-1}Y^{i-1} X^{i-1} X_{i+1}^n|X_i )_{\rho}
&=0 \,.
\end{align}
Hence, 
\begin{align}
 I(A_i C_i Y_i;B^n|A^{i-1} C^{i-1}Y^{i-1} X^n )_{\rho}%
&= I(A_i C_i Y_i;A^{i-1} C^{i-1}Y^{i-1} X^{i-1} X_{i+1}^n B^n| X_i )_{\rho}
\nonumber\\
&\geq 
I(A_i C_i ;B_i| X_i Y_i)_{\rho}
\,.
\label{Equation: Converse Broadcast IID}
\end{align}

Now, define a random variable $J\sim \text{Uniform}[n]$. Let
 $\sigma_{JXYABC}=\sum_{j=1}^n \ketbra{j}_J\otimes \rho_{X_jY_jA_jB_jC_j}$. Thus, by \eqref{Equation: Converse Broadcast Dim(H_M)}, \eqref{Equation: Converse Broadcast Chain rule} and \eqref{Equation: Converse Broadcast IID},
\begin{align}
Q_{1\to 2}&\geq 
\frac{1}{2n}\sum_{i=1}^n I(A_iC_i;B_i|X_i Y_i)_{\rho}
\nonumber\\
&
= \frac{1}{2} I(A_JC_J;B_J|X_JY_J J )_{\sigma}
\,.
\label{Equation: Converse Broadcast Time sharing}
\end{align}

Since Alice begins with preparing a memoryless state, the systems $X_J Y_J A_J C_J $ and $J$ are uncorrelated, as
$\sigma_{JX_J Y_J A_J C_J }=\frac{\identity_J}{n}\otimes \omega_{XY AC}$.
Thus, 
\begin{align}
I(A_J C_J ;B_J|X_JY_JJ)_{\sigma} &=
I(A_J C_J ;B_JJ|X_JY_J)_{\sigma} 
\nonumber\\
& \geq
 I(A_J C_J ;B_J|X_JY_J)_{\sigma}
 \,.
 \label{Equation: Converse Broadcast independent J}
\end{align}
Observe that 
\begin{align}
    \frac{1}{2}\left\Vert \sigma_{X_JY_JA_J B_{J}C_J } - \omega_{XYABC} \right\Vert _{1} \leq \varepsilon_n \,,
\label{Broadcast_Error_Converse_QL_Sigma}
\end{align}
by \eqref{Broadcast_Error_Converse_QL}.
Therefore, by  entropy continuity \cite{AFW_Winter_2016},
\begin{align}
 I(A_J C_J;B_J|X_JY_J)_{\sigma}
&\geq  I(AC;B|XY )_{\omega} - \beta_n
\nonumber\\
&=  2H(B|XY)_{\omega} - \beta_n\,,
\label{Equation: Converse Broadcast AFW}
\end{align}
where $\beta_n=4\varepsilon_n \log \text{dim}(\mathcal{H}_{B})+2(1+\varepsilon_n)h(\frac{\varepsilon_n}{1+\varepsilon_n})$, which tends to zero as $n\to \infty$.
 Recall that
 Bob's system $B$ and Charlie's side information $Y$ are uncorrelated when conditioned on $X$ (see \eqref{Equation:Charlie_Non_Signaling}). Therefore,
\begin{align}
    H(B|XY)_{\omega} =H(B|X)_{\omega} \,.
    \label{Equation: Broadcast 
Uncorrelated Y and B}
\end{align}
Hence, the bound on $Q_{1\to 2}$ follows from \eqref{Equation: Converse Broadcast Time sharing}, \eqref{Equation: Converse Broadcast independent J}, and \eqref{Equation: Converse Broadcast AFW}.
The bound on $Q_{1\to 3}$ is shown in the same manner.
This concludes the proof for Theorem~\ref{Theorem: Broadcast - Quantum links}.
\qed

\section{Summary and Discussion}
\label{Section: Discussion}

\subsection{Summary}
We have introduced the notion of empirical coordination for quantum correlations. %
 Quantum mechanics enables the calculation of probabilities for experimental outcomes, emphasizing statistical averages rather than detailed descriptions of individual events. 
Empirical coordination is thus a natural framework for quantum systems.
 Focusing on the cascade network, we established   the optimal coordination rates,  
indicating the minimal resources for the empirical simulation of a  quantum state. 
As we consider a network with classical communication links, superposition cannot be maintained,   hence the quantum correlations are separable. This precludes entanglement. 
    We have shown that providing the users with shared randomness, before communication begins, does not affect the optimal  rates for empirical coordination
    (see Theorem~\ref{Theorem: CR theorem}). We began with the rate characterization for the basic two-node network (Theorem~\ref{Theorem: Two nodes}), and then generalized to a cascade network (Theorem~\ref{Theorem: Cascade}).  The special case of a network with an isolated node was addressed as well (see Corollary~\ref{Corrolary: Isolated node}).  The results   generalize to other networks as our analysis includes a generic achievability scheme (see Lemma~\ref{Lemma: Generic lemma}) .
    Nonetheless, we do not claim to have solved all coordination scenarios or network topologies. 

 Next, we  discuss the consequences of our results for  quantum cooperative games.

\subsection{Games with quantum actions} %
In many cooperative games, the payoff is associated with the correlation between the players. %
In the \textit{penny matching game}, as introduced by Gossner et al. \cite{gossner2003online}, 
Alice receives a classical sequence $x^n$ from an i.i.d source, thereafter Alice and Bob produce sequences $a^n$ and $b^n$ that should be close to one another and to $x^n$ as well. 
In other words,
Alice and Bob try to guess the source sequence one bit at a time. They gain a point for every bit they both guess correctly. Alice's action $a^n$ is referred to as a guess, even though she knows the original source sequence $x^n$.  As it turns out,  an optimal strategy could let Alice guess wrong, i.e., $a_i\neq x_i$, for some of the time \cite{gossner2003online}. 
Cuff and Zhao \cite{cuff2011coordination} analyzed a generalized version of the game through the classical two-node network.
Here, we introduce a quantum version of the game.

Suppose that Alice receives a classical sequence $x^n$ from an i.i.d source $p_X$, as depicted in the two-node network \ref{Figure: Two nodes}.
The quantum encoding of each user is viewed as the actions \cite{flitney2002introduction}. 
The game is specified by a payoff map 
 \begin{align}
G:\Delta(\mathcal{H}_A\otimes\mathcal{H}_B)\to [0,\infty)
\,.
\end{align}
 Given a joint strategy $\omega_{AB}$, the payoff to Alice and Bob is $G(\omega_{AB})$.
 
Suppose that %
Alice  uses an empirical coordination code and send $nR_{1 \shortrightarrow 2}$ bits to Bob. 
Furthermore, let $\mathrm{SEP}(\gamma)$ be the set of all separable strategies 
 $\omega_{AB}$ for which Alice and Bob receive a payoff
 $\gamma=G(\omega_{AB})$.
Alice and Bob can then reach an average payoff $\gamma\geq 0$ asymptotically,
if and only if Alice can send a message to Bob at rate 
$R_{1 \shortrightarrow 2}>C_{\text{2-node}}(\omega)$ for some $\omega_{AB}\in \mathrm{SEP}(\gamma)$.
The optimal rate $C_{\text{2-node}}(\omega)$ is characterized by Theorem~\ref{Theorem: Two nodes}.

\subsection{Strong coordination vs. empirical coordination}
\label{Subsection:Strong_vs_Empirical}
In analogy to the classical framework, we separate between
two types of coordination tasks; strong coordination and empirical coordination. 

\subsubsection{Strong coordination}
\label{section:Quantum_Strong}
In the classical setting,
strong coordination means that
 a statistician cannot reliably distinguish  between the constructed 
sequence of actions $X_1^n,\ldots,X_K^n$,  and random samples from the desired
distribution \cite{cuff2010coordination}.
This requires
 the joint distribution  $p_{X_1^n \dots X_K^n}$
 that the code induces  to be arbitrarily close to the desired source 
 $\pi\equiv \pi_{X_1 \dots X_K}$ in total variation distance. That is, strong coordination is achieved if there exists a code sequence such that 
\begin{align}
    \lim_{n\to \infty} \norm{ p_{X_1^n\dots X_K^n}-\pi^n}_1=0
    \,,
\label{Classical_strong}
\end{align}
where $\pi^n$ denotes the i.i.d. distribution corresponding to the desired source.

Consider a network of $K$ quantum nodes, 
where the users %
have access to classical (or quantum) communication links with limited rates $R_{k,l}$ ($Q_{k,l}$, respectively), and may share common randomness (CR).
We say that
strong coordination is achieved if  there exists a code sequence %
such that the joint state $\rho_{A_1^n\dots A_K^n}$ that is  the code induces converges to the desired state, i.e., %
\begin{align}
   \lim_{n\to \infty} %
   \norm{ \rho_{A_1^n\dots A_K^n}-\omega^{\otimes n}}_1=0 \,,
    \label{strong_norm}
\end{align}
where $\omega\equiv \omega_{A_1\dots A_K}$  is  the desired state. %
In our previous work \cite{NaturPereg:24c2,natur2025quantum}, we have considered strong coordination
for quantum networks. %

\subsubsection{Empirical coordination}
\label{section:Classical_Empirical}
In the classical description, empirical coordination uses network
communication in order to construct a sequence of actions that have an
empirical joint distribution closely matching the desired distribution \cite{cuff2010coordination}.
In this case, the error criterion sets a weaker requirement, given in terms of the joint \emph{type}, i.e., the  empirical distribution of the actions in the network.
Formally, 
the requirement for
empirical coordination is that for every $\varepsilon>0$,
\begin{align}
    \lim_{n\to \infty} 
    \Pr \left(
    \norm{\hat{P}_{X_1^n  \dots  X_K^n}-\pi}_1
    \geq \varepsilon
    \right)=0
    \,,
\label{Classical_empirical}
\end{align}
where 
$X_1^n, \dots, X_K^n$ are the encoded actions, and the probability is computed with respect to the CR distribution.

We say that
empirical coordination is achieved in a quantum coordination network if  there exists a sequence of coordination codes of length $n$,  such that the time-average state $\frac{1}{n}\sum_{i=1}^n\rho_{A_{1}(i)  \dots A_{K}(i) }$ that is induced by the code converges in probability to the desired source $\omega_{A_1\dots A_K}$, i.e., %
\begin{align}
\lim_{n\to \infty} 
    \Pr \left(
 \norm{\frac{1}{n}\sum_{i=1}^n\rho_{A_{1}(i)  \dots A_{K}(i) } -\omega}_1
    \geq \varepsilon
    \right)=0
    \,,
    \label{empirical_quantum}
\end{align}
where $\omega\equiv \omega_{A_1\dots A_K}$  is  the desired state,  and the probability is computed with respect to the CR distribution.
We note that the quantum definition  differs in nature from the classical one (c.f. \eqref{Classical_empirical} and \eqref{empirical_quantum}).

\begin{remark}
To see that strong coordination is indeed a stronger condition,  note that by trace monotonicity, strong coordination implies 
$ %
   \norm{ \rho_{A_{1}(i)  \dots A_{K}(i) }-\omega}_1\to 0 $ as $n\to\infty$, for every $i\in [n]$. Hence, by the triangle inequality, 
   \begin{align} 
   \norm{\frac{1}{n}\sum_{i=1}^n\rho_{A_{1}(i)  \dots A_{K}(i) } -\omega}_1\leq \frac{1}{n}\sum_{i=1}^n\norm{\rho_{A_{1}(i)  \dots A_{K}(i) } -\omega}_1
   \end{align}
    which also tends to zero as $n\to\infty$.
\end{remark}

We have discussed  the justification and the physical interpretation of our coordination criterion in
Subsection~\ref{Subsection: Justification}. 
Consider an observable represented by an Hermitian operator $\hat{O}$ on $\mathcal{H}_{A_1}\otimes\cdots \mathcal{H}_{A_K}$. In practice, statistics are collected by performing measurements on $n$ systems
$(A_{1}(i) ,\ldots,A_{K}(i)  \,:\; i\in [n])$.
The expected value of the observable in the $i$th measurement is thus,
\begin{align}
\langle \hat{O} \rangle_i&=
\trace\left[ \hat{O}\cdot \rho_{A_{1}(i)  \ldots A_{K}(i) } \right]
\end{align}
for $i\in [n]$.
Roughly speaking, our coordination criterion guarantees that the empirical average is close to the expected value with respect to a desired state, i.e., 
\begin{align}
\frac{1}{n}\sum_{i=1}^n\langle \hat{O} \rangle_i&=
\trace\left[ \hat{O} \cdot \left( \frac{1}{n}\sum_{i=1}^n \rho_{A_{1}(i) \ldots A_{K}(i) } \right) \right]
 \nonumber\\
&\approx
\trace\left[ \hat{O}\cdot   \omega_{A_{1}\ldots A_{K} } \right] 
\,,
\end{align}
 with  high probability.

\subsection{Common randomness does not help}
\label{Section: CR}
We have shown that CR does not improve the coordination capacity. That is, if $R_{k\shortrightarrow l}$ is achievable with CR,  it is also achievable without CR. We provide an intuitive explanation below. 
Suppose  we use a coding scheme where the CR element is a sequence  $U^n$, drawn from a memoryless source $p_U$ over $\mathcal{U}$, and each user  encodes by a collection of maps $\{\mathcal{E}^{(u)}\}$, taking
 $u=U_i$ at time $i$.
Then, this CR-assisted coding scheme can be replaced with a code based on a fixed agreed upon sequence $\tilde{u}^n$ of type $\hat{P}_{\tilde{u}^n}\approx p_U$. %

Since our coding scheme uses binning and not an encoder of the form $\mathcal{E}^{(u_i)}$, the description above is only a rough explanation to gain intuition.

\subsection{Applications}

Recent advances in machine-to-machine communication \cite{mylonakis2020remote} and the Internet of Things (IoT) \cite{torres2023message} have  raised interest in %
networks with various topologies %
\cite{he2020internet}. These network topologies are relevant for various applications, such as %
 distributed computing \cite{borcea2002cooperative}, autonomous vehicles \cite{ahangar2021survey}, 
embedded sensors \cite{stankovic2003real}, players in a cooperative game \cite{cuff2011coordination}, 
and quantum-enhanced IoT  \cite{%
burenkov2021practical,granelli2022novel}.
Coordination  with classical links is motivated by quantum-enhanced IoT networks in which the communication links are classical \cite{notzel2020entanglement,9232550,burenkov2021practical,granelli2022novel}.
 The  problem at hand is to find the optimal transmission rates %
required in order to establish a desired correlation.  
Empirical coordination also plays a role in quantum data compression \cite{horodecki1998limits,barnum2001quantum,soljanin2002compressing}.
The optimal compression rate for a quantum source of pure states was first established by Schumacher \cite{schumacher1995quantum} for a quantum source of pure states (see also \cite{jozsa1994new,barnum1996general}).
Empirical coordination is thus a natural framework for quantum systems.

Empirical coordination also plays a role in quantum data compression \cite{%
soljanin2002compressing}.
Barnum et al. \cite{barnum2001quantum} addressed a source of commuting density operators, and 
Kramer and Savari \cite{kramer2001quantum} developed a 
rate-distortion theory that unifies the visible and blind 
approaches (cf. \cite{dur2001visible} and \cite{horodecki1998limits}). Khanian and Winter have recently solved the general problem of a %
quantum source of mixed states %
(see also \cite{horodecki1998limits,horodecki1999towards,koashi2001compressibility,koashi2002operations,hayashi2006optimal,khanian2020quantum,khanian2022strong}). 
Rate distortion can be viewed as a special case of empirical coordination.

\subsection{Future directions} 
In another work by the authors \cite{NaturPereg_E_arXiv,natur2025quantum}, we have also considered strong coordination in a network with quantum links. This allows for the generation of multipartite entanglement and is closely related to tasks such as quantum  channel/source simulation \cite{%
 berta2013entanglement,
 BennettDevetakHarrowShorWinter:14p,pirandola2018theory,
wilde2018entanglement,%
cheng2023quantum,state_generation_using_correlated_resource_2023,salehi2024quantum},
 state merging \cite{bjelakovic2013universal,horodecki2007quantum},
 state redistribution \cite{Yard_Devetak_2009,%
 berta2016smooth%
 }, zero-communication state transformation \cite{george2023revisiting,george2024reexamination},
 entanglement dilution \cite{hayden2003communication,harrow2004tight,kumagai2013entanglement,
 Salek_2022_Winter},
 randomness extraction \cite{6670761,
 tahmasbi2020steganography},
 source coding \cite{goldfeld2014ahlswede,kramer2001quantum,Compressing_mixed_state_sources_2002, Classical_broadcast_cooperation_2016,
Quantum_Classical_Source_Coding2023,
Rate_limited_source_coding_2023},
 and many others. 
An interesting avenue for future research is to study empirical coordination in such networks.
There are many other
coordination scenarios and network topologies that could be studied further, e.g., empirical coordination with entanglement assistance. 
Other interesting directions include the one-shot setting ($n=1$) and coordination with 
 two-way communication.

\section*{Acknowledgements}
The authors would like to thank Ian George (National University of Singapore), Eric Chitambar (University of Illinois at Urbana-Champaign), and
Marius Junge (University of Illinois at Urbana-Champaign)
for useful discussions during the conference ``Beyond IID in Information Theory," held at the University of Illinois Urbana-Champaign from July 29 to August 2, 2024, supported by NSF Grant n. 2409823.

 H. Natur and U. Pereg were supported by  Israel Science Foundation (ISF), Grants n. 939/23 and 2691/23, German-Israeli Project Cooperation (DIP) n.
2032991, Ollendorff Minerva Center (OMC) of the Technion n. 86160946, and  Nevet Program of the Helen Diller Quantum Center at the Technion n. 	2033613.
U. Pereg was also supported by the Israel VATAT Junior Faculty Program for Quantum Science and Technology  n. 86636903, and the Chaya Career Advancement Chair through Grant n. 8776026.

\bibliography{references}

\end{document}